\documentstyle[prb,aps,multicol,epsfig,amsmath,amssymb,exscale]{revtex}

\renewcommand{\c}[1]{{c^{\rule{0ex}{1ex}}_{#1}}}
\newcommand{\ck}[1]{{c^\dagger_{#1}}}
\newcommand{\cs}[1]{{\c{\vec{#1},\sigma}}}
\newcommand{\cks}[1]{{\ck{\vec{#1},\sigma}}}
\newcommand{\up}{{\uparrow}}
\newcommand{\dn}{{\downarrow}}

\renewcommand{\d}[1]{{\Delta^{\rule{0ex}{1ex}}_{#1}}}
\newcommand{\dk}[1]{{\Delta^\dagger_{#1}}}
\newcommand{\lk}{{\langle}}
\newcommand{\rk}{{\rangle}}

\newcommand{\cc}[1]{{\tilde{c}^{\rule{0ex}{1ex}}_{#1}}}
\newcommand{\cck}[1]{{\tilde{c}^\dagger_{#1}}}
\newcommand{\gc}[1]{{\gamma^{c\rule{0ex}{1ex}}_{#1}}}
\newcommand{\gck}[1]{{\gamma^{c\dagger}_{#1}}}
\newcommand{\gv}[1]{{\gamma^{v\rule{0ex}{1ex}}_{#1}}}
\newcommand{\gvk}[1]{{\gamma^{v\dagger}_{#1}}}
\newcommand{\g}{{{\mathcal{G}}^{\scriptscriptstyle HF}}}
\newcommand{\gs}{{\widetilde{{\mathcal{G}}\,}^{\scriptscriptstyle HF}}}
\newcommand{\overl}[1]{\overline{#1}}

\long\def\changed#1{#1}
\long\def\changede#1{#1}

\DeclareMathOperator{\sign}{sign}

\begin{document}

\title{
$t-U-W$ Model of a $d_{x^2-y^2}$ Superconductor in the Proximity of
 an AF Mott Insulator:
Diagrammatic Studies vs. Quantum-Monte-Carlo Simulations
}

\author{T.~Eckl, E.~Arrigoni and W.~Hanke \\
\textsl{\small Institut f\"ur Theoretische Physik, Universit\"at 
W\"urzburg, 
Am Hubland, D-97074 W\"urzburg, Germany}\\[2ex]
 F.~F.~Assaad\\
\textsl{\small Institut f\"ur Theoretische Physik, Universit\"at 
Stuttgart, 
Pfaffenwaldring 57, D-70550 Stuttgart, Germany}}

\date{\today}

\maketitle

\begin{abstract}
We examine the competition and relationship between an antiferromagnetic 
(AF) Mott insulating state
and a $d_{x^2-y^2}$ superconducting (SC) state in two dimensions using 
semi-analytical, i.~e.~diagrammatic
calculations of the $t-U-W$ model. The AF Mott insulator is described by 
the ground state
of the half-filled Hubbard model on a square lattice with on-site Coulomb 
repulsion $U$ and nearest 
neighbor single-particle hopping $t$. To this model, an extra term $W$ is 
added, which depends upon 
the square of the single-particle nearest-neighbor hopping. Staying at 
half-band filling and a constant
value of $U$, it was previously shown with Quantum-Monte-Carlo (QMC)
simulations that one can generate a quantum transition as a function
of the coupling strength, $W$, between an AF Mott insulating state
and a $d_{x^2-y^2}$ SC state.
Here we complement these earlier QMC simulations with physically more 
transparent diagrammatic calculations.
We start with a standard Hartree-Fock (HF) calculation to capture the 
``high-energy'' physics of the $t-U-W$ model.
Then, we derive and solve the Bethe-Salpeter equation, i.~e.~we account 
for fluctuation effects within 
the time-dependent HF or generalized RPA scheme. Spin- and 
charge-susceptibility as well as the
effective interaction vertex are calculated and systematically compared 
with QMC data.
Finally, the corresponding BCS gap equation obtained for this effective 
interaction is solved.
\end{abstract}

\section{Introduction }

One salient aspect of the high-$T_c$ materials is the vicinity of two, at
first sight rather different, states of matter, superconductivity (SC) and
 antiferromagnetism 
(AF) in their phase diagram.
The transition between the undoped (AF) system at half-filling and the 
SC phase is driven by doping with mobile
holes.
In most of the materials, this transition is not direct, and a disordered
``spin-glass'' phase occurs in between. However, it has been argued
that the ``clean'' material would display a direct transition from 
AF to SC phases, and that the spin-glass phase occurs
due to the high sensitivity to impurity disorder in the vicinity
of the phase transition.

A direct transition from an insulating into a SC phase
in a quasi-two-dimensional system (such as the high-$T_c$ materials)
is a very interesting, yet insufficiently understood issue.
In fact, it is not clear whether this transition is second order down
to zero temperature, and thus is related to a quantum critical point,
or whether there is a finite-temperature classical bicritical point.
In the framework of 
the projected SO(5)-theory of
high-$T_c$ superconductivity~\cite{pso5},
it has been suggested that the AF and the SC phases may indeed coexist
in some portion of the temperature versus doping phase diagram.
Another open question is the nature, i.~e.~the universality
class of this transition.
For example,
it has been suggested that this transition may be controlled by an
SO(5)-symmetric fixed point~\cite{mu.na.99}.
SO(5) symmetry is thus restored in the long-wavelength limit~\cite{so5.97},
and AF and SC can be
described in terms of a unique superspin vector~\cite{zhan.97} in the
vicinity of the critical point.

Many efforts have been directed towards studying the AF-SC transition 
in strongly correlated lattice models
by numerical techniques such as Quantum-Monte-Carlo (QMC) simulations.
As a relevant model, the Hubbard model, is widely accepted for the
description of salient features of high-$T_c$ materials.
Unfortunately, it is quite difficult to study large enough
Hubbard-model systems  
by  QMC, due to the occurrence of the minus-sign
problem at finite doping.
The numerical problem can, in principle, be cured, if one can drive the
AF-SC transition by means of a parameter, alternative to the doping,
which conserves particle-hole symmetry and, therefore, avoids the
tedious minus-sign  problem. 
This idea was followed through by Assaad, Imada, and Scalapino 
(AIS)~\cite{tuw1}
in terms of their so-called $t-U-W$ model. It rests on adding an 
interaction term $W$,
which depends on the square of the nearest-neighbor hopping. 
This $W$ term can be obtained from a Su-Schrieffer-Heeger 
type of electron-phonon interaction in the antiadiabatic limit \cite{ssh}.
In QMC simulations \cite{tuw1,tuw2,tuw3,tuw4,tuw5} this $t-U-W$ model 
exhibits a 
transition from an antiferromagnet to a d-wave superconductor at 
half-filling and at a 
critical value of the interaction $W_c\approx 0.3 t$ ($U=4t$, $T=0K$).
The QMC data of AIS supports the picture of a continuous quantum phase 
transition in the sense
that the magnetization vanishes continuously at the critical point.
The disadvantage of the $t-U-W$ model is that  the bandwidth grows 
substantially 
with $W$.
Therefore, one of us \cite{tuwph} suggested to introduce a phase factor
in the $W$-term which has a d-wave like symmetry.
Although this latter model solves the problem of the bandwidth, the 
existence of a 
phase transition to a d-wave superconductor remains open.

While QMC calculations provide an essentially exact description of the 
properties of the model, semi-analytical, i.~e.~diagrammatic,
calculations 
allow for  a more direct understanding of the 
processes which are responsible for a given phenomenon.
For this reason in this paper, we carry out a systematic diagrammatic study
of the $t-U-W$ model.

We first consider in Sec.~\ref{hartree} the simple
Hartree-Fock level, which, due to the complexity of the interaction
terms, allows for different broken symmetry phases. However, a careful 
comparison of the energies of these phases shows that the
antiferromagnetic phase is always the stablest one, even for very large 
values of $W$. This holds for both versions of the 
$t-U-W$ model which are considered, i.~e.~with and without phase
factors. 
Moreover, the only allowed superconducting solution in the simple
$t-U-W$ model has an $s$-wave symmetry, while $d$-wave symmetry is not 
allowed. These mean-field results are in strong contrast with the QMC 
calculations, which
predict a transition to a $d$-wave SC state at some finite
$W$~\cite{tuw1}.
On the other hand, in the AF region our mean-field results are in very
good accord with QMC, in particular concerning single-particle
dispersions, as shown in Sec.~\ref{band}.

\changede{The fact that the transition to the superconducting state does not
come out correctly 
 is of no surprise 
within an Hartree-Fock approximation.}
Indeed
the relevant transition to the d-wave SC state is dominantly
driven by an effective attractive interaction mediated by spin fluctuation,
as is the case for the Hubbard model~\cite{swz,hanfre}. For this reason, one 
has to consider the effect of spin fluctuations beyond the
Hartree-Fock level, in order to reach the SC state. 
This is done in Sec.~\ref{rpa}, where we carry out a complete RPA
summation of all particle-hole diagrams (both bubbles and ladders),
with Hartree-Fock Green's functions,
in order to obtain the frequency- and momentum-dependent
spin and charge susceptibility.
The solution of the corresponding Bethe-Salpeter equation
 is technically quite difficult to achieve and significantly more demanding
than the standard case of the simple Hubbard model~\cite{swz,hanfre}. This is 
due to the finite extension of 
the interaction, as well as its dependence on all (three) momenta, and
not on the momentum transfer only. By changing to a mixed real-space
momentum-space representation, we demonstrate that it can be reduced,
for generic momenta, to the inversion of a $52\times52$ matrix.

Next, in some analogy to the RPA-analysis of the $t-U$ Hubbard model by
Schrieffer, Wen, and Zhang~\cite{swz}, we derive the effective
two-particle interaction vertex in the static limit,  
and solve the associated BCS equation.
As for the simple Hubbard model studied in Ref.~\onlinecite{swz}, 
the $d$-wave solution turns out to be the only stable one, 
in accordance with QMC results. 
On the other hand, as expected, 
the $s$-wave solution is unstable, due to
the strong on-site Hubbard repulsion $U$. 
In the d-wave phase,
we obtain a decreasing superconducting gap as a
function of $W$, in spite of the fact that the attraction between the
quasiparticles should be increased by $W$. 
This is due, on the one hand, to the approximation of taking
an energy cutoff for the
effective interaction, which has been chosen to be of the order of 
the AF gap, which, in turn, decreases with increasing $W$.
On the other hand,
the reduction of the density of states at the Fermi
level, which is 
related to  the broadening of the bands produced by $W$,
contributes in reducing the superconducting gap.

Our paper is organized as follows. In Sec.~\ref{model}
the $t-U-W$ model with and without phase factors is introduced, and 
briefly summarized
In Sec.~\ref{hartree}, we carry out the
Hartree-Fock mean-field study of the antiferromagnetic phase.
We discuss the HF results and
compare them with QMC calculations.
In Sec.~\ref{rpa} we derive and solve the Bethe-Salpeter equation,
i.~e.~we account for fluctuation effects in the time-dependent
HF or generalized RPA scheme. We obtain 
spin and charge susceptibilities, as well as the effective
interaction vertex.
In Sec.~\ref{bcs}, we write down and solve the BCS gap equation, 
obtained from this effective interaction within a static approximation.
Finally, we present our conclusions in Sec.~\ref{summary},
partly based on detailed comparisons with QMC data.

\section{Model }
\label{model}

The Hamiltonian of the $t-U-W$ model is given by \cite{tuw1}
\begin{equation}\label{1}
{\mathcal{H}}=-\frac{t}{2} \sum_{\vec{i}}K_{\vec{i}}+U\sum_{\vec{i}}
(n_{\vec{i},\up}-{\textstyle \frac{1}{2}})
(n_{\vec{i},\dn}-{\textstyle\frac{1}{2}})-W \sum_{\vec{i}} 
\widetilde{K}_{\vec{i}}^2
\end{equation}
with the hopping kinetic energy
\begin{equation}\label{2}
K_{\vec{i}}=\sum_{\sigma,\vec{\delta}} (\cks{i} \c{\vec{i}+\vec{\delta},
\sigma} +
\ck{\vec{i}+\vec{\delta},\sigma} \cs{i})
\end{equation}
and
\begin{equation}\label{3}
\widetilde{K}_{\vec{i}}=\sum_{\sigma,\vec{\delta}} f(\vec{\delta})
(\cks{i} \c{\vec{i}+\vec{\delta},\sigma} +
\ck{\vec{i}+\vec{\delta},\sigma} \cs{i}) \;,
\end{equation}
where in its original formulation $ f(\vec{\delta})=1$.
As mentioned in the Introduction, this  model 
was introduced by Assaad, Imada, and Scalapino~\cite{tuw1}, in order 
to study the
 antiferromagnetic-superconducting transition at half-filling.
As stressed by these authors \cite{tuw1,tuw2,tuw4,tuw5}, 
the particular choice of $W$ was mainly motivated
formally as a means of introducing the desired quantum transition
from the insulating (AF) to the SC state. 
The choice of the interaction also guaranteed
that no fermion-sign problem was encountered in the QMC simulations at
half-filling. 
While there are various approximate ways to physically justify the form 
of the microscopic Hamiltonian, it runs
into problems when directly compared to the high-$T_c$ cuprates.
One of the problems of this model is the presence of unphysically broad
 bands, which also cause problems in the numerical QMC evaluation.
For this reason,
one of us \cite{tuwph} suggested to introduce
 in the $\widetilde{K}_{\vec{i}}\,$--term of Eq.~(\ref{3}) a d-wave like 
phase factor of the form
\begin{equation}\label{4}
f(\vec{\delta})=\begin{cases}+1&\text{for\quad}\vec{\delta}=
\pm\vec{a}_x \\ -1&\text{for\quad}\vec{\delta}=
\pm\vec{a}_y \end{cases}.
\end{equation}
The phase factor restores the correct width of the single--particle
bands at half-filling. 
Unfortunately, however no
transition to a superconductor in QMC simulations \cite{tuwph} has
been observed so far.

The $W$ term contains four different processes (see e.g. 
Ref.~\onlinecite{tuw2}), among them single--particle 
terms that renormalize the chemical potential and permit single--particle 
hopping between
second and third nearest-neighbor sites as well as singlet and triplet 
scattering terms.
These terms are not expected
 to be relevant for the low-energy physics \cite{tuw2}.
However, the most interesting term for the quantum phase transition is
the fourth  term which 
generates singlet pair-hopping and produces an antiferromagnetic 
exchange interaction, i.~e.
\begin{equation}\label{4a}
{\mathcal{H}}_W^{(4)}=-2W\sum_{\vec{i},\vec{\delta},\vec{\delta}^\prime}
f(\vec{\delta})
f(\vec{\delta}^\prime)\dk{\vec{i},\vec{\delta}^\prime}\d{\vec{i},
\vec{\delta}},
\end{equation}
where $\dk{\vec{i},\vec{\delta}}=(\ck{\vec{i},\up}\ck{\vec{i}+
\vec{\delta},\dn}-
\ck{\vec{i},\dn}\ck{\vec{i}+\vec{\delta},\up})/\sqrt{2}$. For 
$\vec{\delta}=\vec{\delta}^\prime$ the terms in
${\mathcal{H}}_W^{(4)}$ contribute to the exchange giving:
\begin{equation}\label{4b}
2W\sum_{\vec{i},\vec{\delta}}(\vec{S}_{\vec{i}}\cdot\vec{S}_{\vec{i}+
\vec{\delta}}-
{\textstyle\frac{1}{4}}n_{\vec{i}}n_{\vec{i}+\vec{\delta}}).
\end{equation}

\section{Hartree-Fock Calculations}
\label{hartree}

\label{hfresults}

\begin{figure}[h]
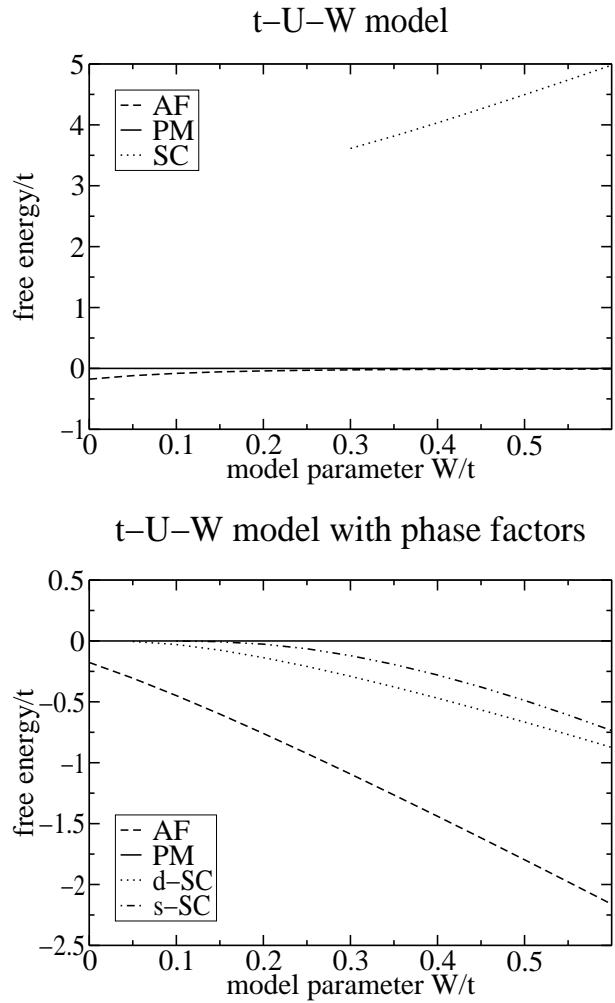

\begin{center}
\epsfig{file=fig1a.eps,width=8cm}
\end{center}
\begin{center}
\epsfig{file=fig1b.eps,width=8cm}
\end{center}
\caption[free energy, $t-U-W$ model]{Free energy per lattice site of 
the different phases of the simple 
$t-U-W$ model (top) and the $t-U-W$ model with phase factors (bottom). 
Plotted is the difference 
to the paramagnetic energy ($U=4t$, $T=0K$, $\mu=0\,t$, lattice size: 
$60\times60$).}
\label{fig1a}
\label{fig1b}
\end{figure}

The details of our HF calculation are given in Appendix \ref{a1}.
After solving the self-consistent equations for the mean-field parameters  
in Eqs.~(\ref{5a}) to (\ref{6}) and (\ref{10a}) to (\ref{11}) (Appendix 
\ref{a1}), 
we arrive at the following results:


Figure \ref{fig1a} displays  the free energy of 
the different phases (antiferromagnetic, superconducting and paramagnetic)
for the simple $t-U-W$ model (top) and the $t-U-W$ model with phase factors 
(bottom) 
as a function of $W$ for fixed $U=4t$ and $T=0K$.
\changed{For the sake of comparison, we only plot the difference to the 
paramagnetic energy.}

In the simple $t-U-W$ model (Fig.~\ref{fig1a}, top) the antiferromagnetic 
solution (AF) is always the most favorable.
However, with increasing $W$ the energy of this solution approaches the 
paramagnetic solution (PM).
The superconducting solution (SC) has a much higher energy than the other 
solutions.
The only possible superconducting solutions have 
$s_1=s_3=0$ (see Appendix \ref{a1}: no on-site pairing, due to $U$)
and $s_2\ne0$ (nearest-neighbor singlet pairing), while for $W\lesssim0.3t$ there 
exists no 
superconducting solution. The superconducting order parameter $s_2$ 
corresponds to an s-wave like
symmetry. 
\changed{The  transition from an antiferromagnet to a 
$d_{x^2-y^2}$-superconductor
observed in QMC simulations is not reproduced
at the mean-field level.}

In the $t-U-W$ model with phase factors (Fig.~\ref{fig1b}, bottom) the 
mean-field 
ground state is also antiferromagnetic (AF).
Here, however, in contrast to the simple $t-U-W$ model, the difference in 
energy with the paramagnetic (PM) solution is
increasing with increasing $W$. As discussed in Appendix A, there also 
exist two different superconducting solutions,
that lie energetically between the antiferromagnetic and the paramagnetic 
solution and evolve continuously
from the paramagnetic solution at $W=0t$.
In the first solution,  $s_1$ and $s_3$ are nonvanishing, while  $s_2=0$ 
(s-wave). The order
 parameter has a s-wave symmetry 
with a superimposed weak modulation of the gap. In the second, 
energetically more favorable, solution
one has $s_1=s_3=0$, and $s_2\ne0$, yielding an order parameter with 
d-wave symmetry.
Notice that also in QMC simulations at half-filling, no transition to a 
superconductor was found
in the $t-U-W$ model with phase factors, in agreement with our results.
However, the mean-field result in Fig.~\ref{fig1b} is promising in 
direction of doping away
from half-filling, where the AF phase is suppressed.

\label{band}

\begin{figure}[h]
\begin{center}
(a)\epsfig{file=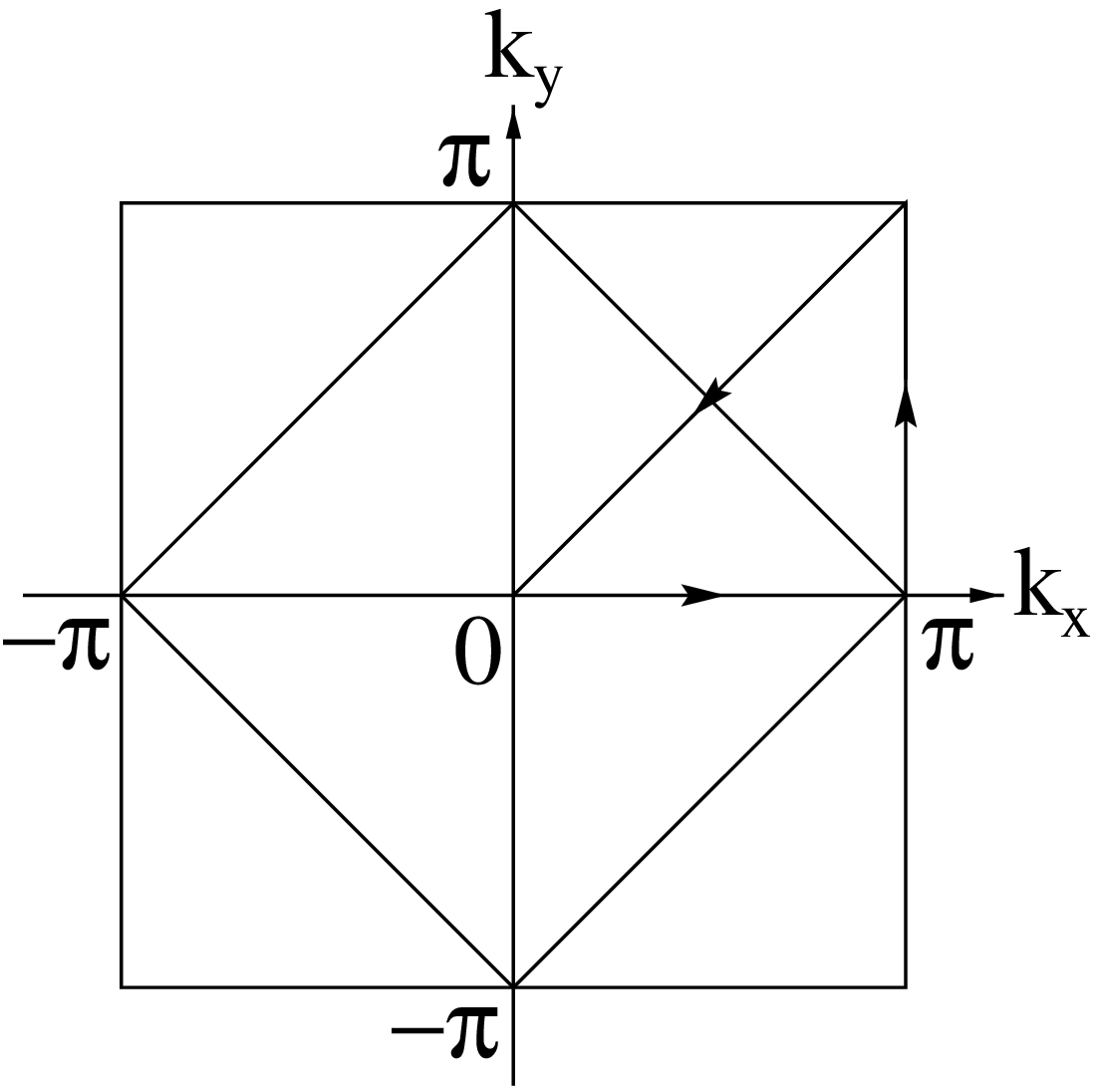,height=3.5cm}
(b)\epsfig{file=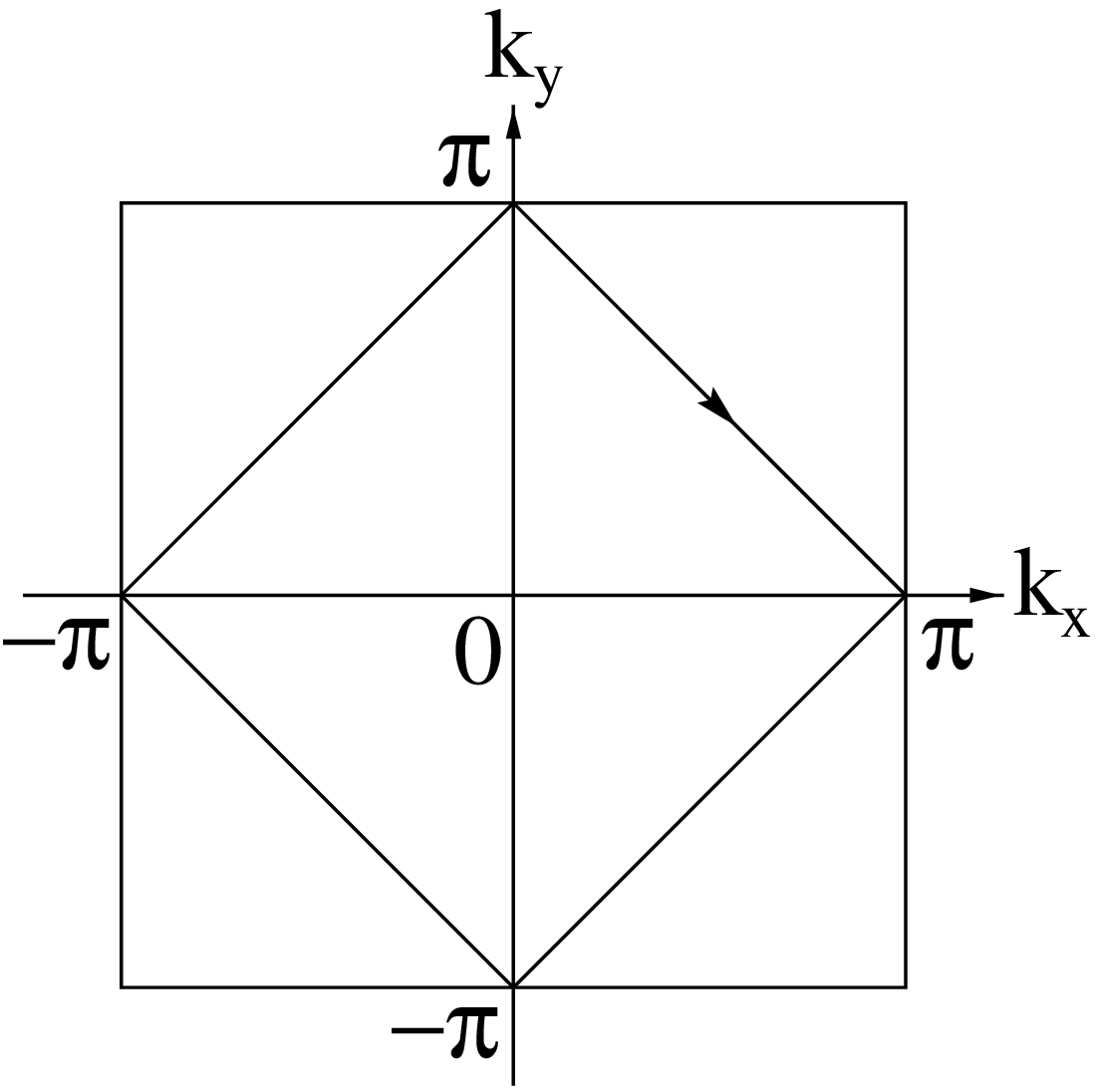,height=3.5cm}
\end{center}
\caption[paths through BZ]{Paths followed through the Brillouin zone to 
plot the energy bands of
Fig.~\ref{fig3a}.}
\label{fig2}
\end{figure}

\begin{figure}[h]
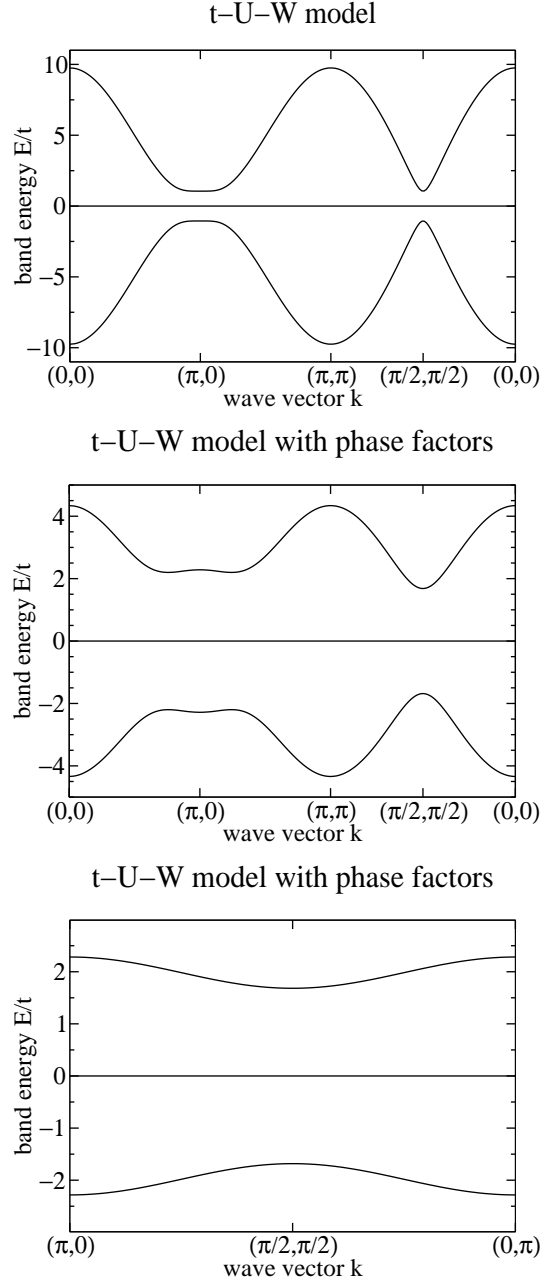

\begin{center}
\epsfig{file=fig3a.eps,width=7cm}
\end{center}
\begin{center}
\epsfig{file=fig3b.eps,width=7cm}
\end{center}
\begin{center}
\epsfig{file=fig3c.eps,width=7cm}
\end{center}
\caption[energy bands]{Energy bands of the simple $t-U-W$ model 
(upper graph, $W=0.15t$)
and of the $t-U-W$ model with phase factors (middle and lower graph, 
$W=0.05t$)
on the paths shown in Figs.~\ref{fig2}(a), for upper and middle graph, 
and \ref{fig2}(b) for lower graph
($U=4t$, $T=0K$, $\mu=0\,t$, lattice size: $100\times100$).}
\label{fig3a}
\label{fig3b}
\label{fig3c}
\end{figure}

The band structure of the antiferromagnetic solutions is evaluated along 
the usual
paths through the Brillouin zone, as shown in Fig.~\ref{fig2}.
Fig.~\ref{fig3a} (top) gives
the bands of the simple $t-U-W$ model for $W=0.15t$.
One can recognize easily that the bands are much wider than in the Hubbard
 model but
their shape is nearly unaltered.
This means that if one would scale the bands by a factor $\frac{1}{3}$, 
they would be
almost identical. The effect of $W$ seems thus to be a mere ``dilatation'' 
of the bands.

In the $t-U-W$ model with phase factors, things look quite different. The 
bands are plotted in Fig.~\ref{fig3b} (middle)
along the path shown in Fig.~\ref{fig2}(a) and in Fig.~\ref{fig3c} (bottom)
 along the path 
shown in Fig.~\ref{fig2}(b) with $W=0.05t$. 
The width of the bands is nearly the same as in the Hubbard model, 
except for the lifting of the
degeneracy along the boundaries of the magnetic Brillouin zone (MBZ).
At $\vec{k}=(\pi,0)$ a kind of double-hump structure can be seen like it 
appears in 
$t$-$t^{\prime}$-$t^{\prime\prime}$ models to describe high-$T_c$ 
superconductors \cite{yin}.
This can be explained as follows: The $W$ term contains also hopping 
processes to
second and third nearest neighbor sites \cite{tuw2,tuwph} which, due to 
the phase factors, 
have the same sign as in the standard fit parameters $t,t^{\prime},
t^{\prime\prime}$ 
which are often used to 
adjust the bands to the experimental data of 
Bi$_2$Sr$_2$CaCu$_2$O$_{8+\delta}$ and YBa$_2$Cu$_3$O$_{7-\delta}$.

\begin{figure}[h]
\begin{center}
\epsfig{file=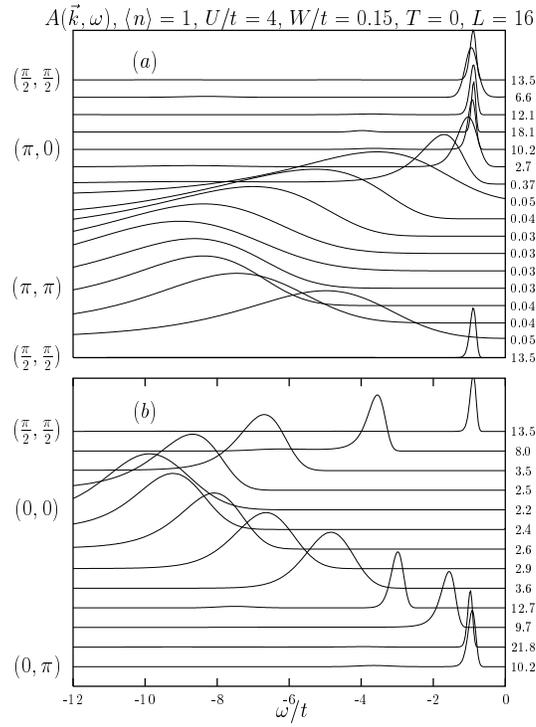,width=7cm}
\end{center}
\caption[]{Spectral weight A($\vec{k}$,$\omega$) of the simple $t-U-W$ 
model, obtained from QMC simulations
($U=4t$, $W=0.15t$, $T=0K$, $\mu=0\,t$) (from \onlinecite{tuw5}).
The considered path in the Brillouin zone is listed
on the left hand side of the figure. We have normalized the raw data
by the factor listed on the right hand side of the figure.
This normalization sets the peak value of $A(\vec{k}, \omega )$ to unity
for all considered $\vec{k}$ vectors.  }
\label{fig4}
\end{figure}

\begin{figure}[h]
\begin{center}
\epsfig{file=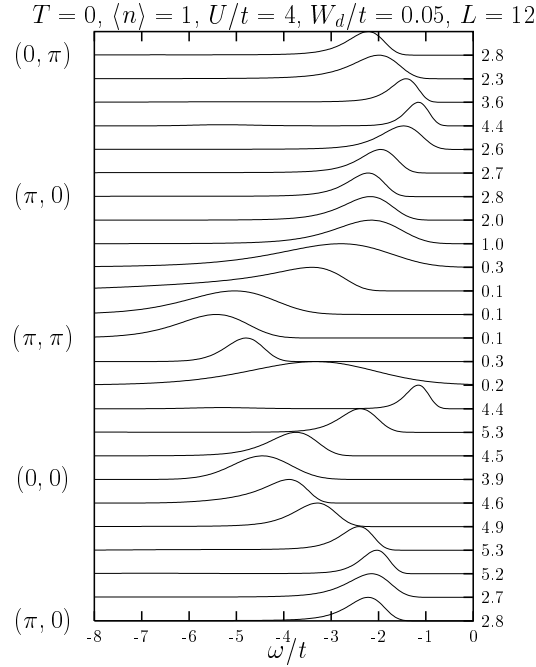,width=7cm}
\end{center}
\caption[]{Spectral weight A($\vec{k}$,$\omega$) of the $t-U-W$ model 
with phase factors, obtained from QMC simulations
($U=4t$, $W=0.05t$, $T=0K$, $\mu=0\,t$).  }
\label{fig5}
\end{figure}

If one compares the antiferromagnetic bands to the QMC data 
\cite{tuw5,tuwph,mgz} 
of the $t-U-W$ model as in Figs.~\ref{fig4} and \ref{fig5},
one gets a very good agreement. The width of the bands as well as the 
antiferromagnetic gap were reproduced
excellently. In addition, the energy bands of the $t-U-W$ model with phase
 factors (Fig.~\ref{fig3b}) show the same
double-hump structure at $\vec{k}=(\pi,0)$ like it is seen in the QMC 
spectral weight $A(\vec{k},\omega)$ 
(Fig.~\ref{fig5}).
They also reproduce well the lift of the degeneracy along the boundaries 
of the MBZ.

\begin{figure}[h]
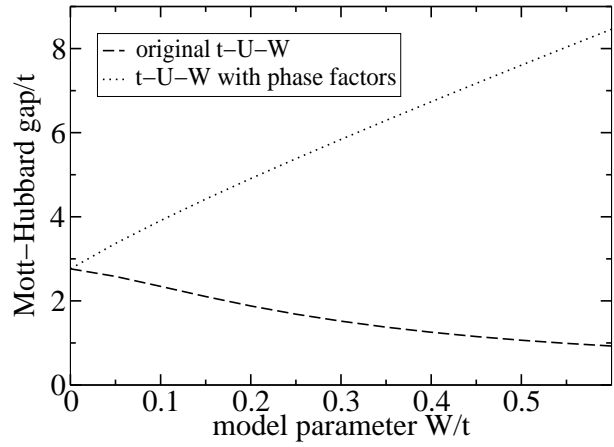

\begin{center}
\epsfig{file=fig6a.eps,width=8cm}
\end{center}
\begin{center}
\epsfig{file=fig6b.eps,width=8cm}
\end{center}
\caption[sublattice magnetization, Mott-Hubbard gap]{Sublattice 
magnetization (top) and 
Mott-Hubbard gap (bottom) of the $t-U-W$ 
model as a function of $W$ ($U=4t$, $T=0K$, $\mu=0\,t$, lattice size: 
$100\times100$).}
\label{fig6a}
\label{fig6b}
\end{figure}

Finally, we want to look at two characteristic features of the 
antiferromagnetic solution:
the sublattice magnetization $m$ and the Mott-Hubbard gap.
The sublattice magnetization defined as
\begin{equation}\label{12}
m=|\lk\ck{\vec{i},\up}\c{\vec{i},\up}\rk-\lk\ck{\vec{i},\dn}
\c{\vec{i},\dn}\rk|\;,
\end{equation}
is plotted
in Fig.~\ref{fig6a} (top) as a function of $W$. As expected from the 
behavior of the free energy (Fig.~\ref{fig1a}),
the sublattice magnetization decreases  with increasing $W$ in the simple 
$t-U-W$ model.
On the other hand, 
 the sublattice magnetization of the $t-U-W$ model with phase factors is 
getting stronger with 
increasing $W$. This is also confirmed by QMC data \cite{tuwph}, which 
show an amplification of
the antiferromagnetic correlations with increasing $W$.

 A similar picture occurs for the antiferromagnetic gap (Fig.~\ref{fig6b},
 bottom). Like the sublattice magnetization,
the Mott-Hubbard gap is decreasing  with increasing $W$ in  the simple
$t-U-W$ model, while
 it increases (nearly linear) with $W$ in the model with phase factors.


In summary, 
the Hartree-Fock calculation gives the antiferromagnetic
solution as ground state for any values of $W$ in both models.
 However,  qualitatively there are remarkable differences between the
 two models with and without phase factors.
A comparison of these results with the QMC data
 shows that the antiferromagnetic phase is described
quite well by the mean-field approximation.

On the other hand, 
the mean-field level is not able to reproduce the transition 
to a $d_{x^2-y^2}$-superconductor 
at $W_c\approx0.3t$ observed in QMC simulations in the simple $t-U-W$ model.
\changede{
This is of no surprise since  results of an Hartree-Fock approximation 
at finite values of the interaction should be taken with due care 
and
cannot give decisive conclusions about the correct phase diagram of a
model without a comparison with  more reliable calculations, such as,
e. g. QMC.
}

Nevertheless, 
for large $W$, for which the AF gap becomes small, 
one would expect the antiferromagnetic solution to become instable
with respect to fluctuations beyond the mean-field level.
This is what we analyze in the next section.

\section{Time-dependent Hartree-Fock (generalized RPA)}
\label{rpa}

As demonstrated in the previous section, the HF mean-field approximation 
is not sufficient to describe
the transition to a $d_{x^2-y^2}$-superconductor occurring in the simple 
$t-U-W$ model
according to QMC simulations.
For that reason, we carried out an improved calculation, including
charge- and spin-density fluctuations.
This has been done by means of a 
time-dependent HF or generalized random phase approximation (RPA), in 
which we summed both
"bubble" and  "ladder" particle-hole diagrams. 
In contrast to the fluctuation exchange approximation (FLEX), the 
Green's functions are not
calculated selfconsistently, but taken over from the Hartree-Fock
results, as it has been done in Ref.~\onlinecite{swz}.

\subsection{Hartree-Fock Correlation Function $L^{{\scriptscriptstyle 0}}$
 and Interaction Vertex 
$\Gamma^{{\scriptscriptstyle 0}}$}

The $2\times2$ antiferromagnetic Hartree-Fock Green's function
can be written as (see Appendix \ref{a1}, Eqs.~(\ref{5}) to (\ref{9})):
\begin{equation}\label{13}
\g(\vec{k},\omega,\sigma)=\begin{pmatrix}
i\omega+\varepsilon(\vec{k})&\sigma\Delta(\vec{k})\\
\sigma\Delta(\vec{k})&i\omega-\varepsilon(\vec{k})
\end{pmatrix}
\frac{1}{-\omega^2-E^2(\vec{k})}.
\end{equation}
With this Green's function we can construct the Hartree-Fock two-particle 
propagator $L^{{\scriptscriptstyle 0}}$:
\begin{equation}\label{14}
L^{{\scriptscriptstyle 0}}\;^{\substack{\scriptscriptstyle m_1\,
m_2\\\scriptscriptstyle m_1^\prime\,m_2^\prime}}
_{\!\!\substack{\scriptscriptstyle \sigma_1\,\sigma_2\\\scriptscriptstyle 
\sigma_1^\prime\,\sigma_2^\prime}}
(\vec{k}_1,\vec{k}_2,\vec{q},\omega_1,\omega_2,\nu)=\delta_{\sigma_1\,
\sigma_2}
\delta_{\sigma_1^\prime\,\sigma_2^\prime}\delta_{\vec{k}_1\,\vec{k}_2}
\delta_{\omega_1\,\omega_2}
\g_{m_2\,m_1}(\vec{k}_1,\omega_1,\sigma_1)\g_{m_1^\prime\,m_2^\prime}
(\vec{k}_1-\vec{q},\omega_1-\nu,\sigma_1^\prime),
\end{equation}
where the set of $m_i$ stand for the indices of the $2\times2$ matrix in 
Eq.~(\ref{13}).
After a unitary transformation with help of the Pauli matrices, i.~e.~
\begin{equation}\label{15}
\widetilde{L}^{{\scriptscriptstyle 0}}=U^\dagger 
L^{{\scriptscriptstyle 0}} U,
\end{equation}
where
\begin{equation}\label{16}
U_{\substack{\sigma_1 \\ \sigma_2}\,\alpha}=\frac{1}{\sqrt{2}}
\sigma^\alpha_{\sigma_1\,\sigma_2}\,,
\quad \alpha=\scriptstyle 0,x,y,z\,,
\end{equation}
we can write the correlation function in the charge-/ spin-channel
representation as:
\begin{equation}\label{17}\begin{split}
&\widetilde{L}_{a\,b}^{{\scriptscriptstyle 0}}\,^{\substack{
\scriptscriptstyle m_1\,m_2\\
\scriptscriptstyle m_1^\prime\,m_2^\prime}}(k_1,k_2,q)=\\&\qquad\qquad\quad
\begin{pmatrix}
\widetilde{L}_{0\,0}^{{\scriptscriptstyle 0}}\,^{\substack{
\scriptscriptstyle m_1\,m_2\\
\scriptscriptstyle m_1^\prime\,m_2^\prime}}(k_1,k_2,q)&0&0&
\widetilde{L}_{0\,z}^{{\scriptscriptstyle 0}}
\,^{\substack{\scriptscriptstyle m_1\,m_2\\\scriptscriptstyle m_1^\prime\,
m_2^\prime}}(k_1,k_2,q)\\
0&\widetilde{L}_{+\,-}^{{\scriptscriptstyle 0}}\,^{\substack{
\scriptscriptstyle m_1\,m_2\\
\scriptscriptstyle m_1^\prime\,m_2^\prime}}(k_1,k_2,q)&0&0\\
0&0&\widetilde{L}_{-\,+}^{{\scriptscriptstyle 0}}\,^{\substack{
\scriptscriptstyle m_1\,m_2\\
\scriptscriptstyle m_1^\prime\,m_2^\prime}}(k_1,k_2,q)&0\\
\widetilde{L}_{z\,0}^{{\scriptscriptstyle 0}}\,^{\substack{
\scriptscriptstyle m_1\,m_2\\
\scriptscriptstyle m_1^\prime\,m_2^\prime}}(k_1,k_2,q)&0&0&
\widetilde{L}_{z\,z}^{{\scriptscriptstyle 0}}\,^{\substack{
\scriptscriptstyle m_1\,m_2\\
\scriptscriptstyle m_1^\prime\,m_2^\prime}}(k_1,k_2,q)
\end{pmatrix}.
\end{split}\end{equation}
Here and in the following: $k=(\vec{k},\omega)$, $q=(\vec{q},\nu)$ and so 
on.

In contrast to the Hubbard model, also  
the non-diagonal elements 
which couple the charge channel to the longitudinal spin channel have to 
be taken into account.
For the following calculations it is also advantageous to transform from 
the representation
\begin{equation}\label{18}
\widetilde{L}_{a\,b}^{{\scriptscriptstyle 0}}\,^{\substack{
\scriptscriptstyle m_1\,m_2\\
\scriptscriptstyle m_1^\prime\,m_2^\prime}}(\vec{k}_1,\vec{k}_2,\vec{q},
\omega_1,\omega_2,\nu)
\text{\qquad with \qquad}\vec{k}_1,\vec{k}_2,\vec{q}\in MBZ
\end{equation}
to the representation
\begin{equation}\label{19}
\overl{L}_{a\,b}^{{\scriptscriptstyle 0}}
(\vec{k}_1,\vec{k}_2,\omega_1,\omega_2;\vec{q}+n\,\vec{Q},\vec{q}+n^\prime
\vec{Q},\nu)
\text{\qquad with \qquad}\vec{k}_1,\vec{k}_2 \in BZ,\quad \vec{q}\in MBZ,
\end{equation}
where $n$, $n^\prime$ take the values $\{0, 1\}$ and $\vec{Q}=(\pi,\pi)$.

In this representation, e.~g., the longitudinal
spin correlation function can be written as a matrix in the indices $n$, 
$n^\prime$:
\begin{equation}\label{20}
\begin{split}
\overl{L}_{z\,z}^{{\scriptscriptstyle 0}}&(k_1,k_2;\vec{q}+n\,\vec{Q},
\vec{q}+n^\prime\vec{Q},\nu)=
\\&\quad\delta_{\omega_1\,\omega_2}
\begin{pmatrix}
\delta_{\vec{k}_1\,\vec{k}_2} \gs_{1\,1}(k_1)\gs_{1\,1}(k_1-q)+\delta_{
\vec{k}_1\,\,\vec{k}_2+\vec{Q}}
\gs_{2\,1}(k_1)\gs_{1\,2}(k_1-q)\quad \qquad \qquad 0 \quad \qquad\qquad\\
\quad\qquad \qquad 0 \quad \qquad\qquad \delta_{\vec{k}_1\,\vec{k}_2} 
\gs_{1\,1}(k_1)\gs_{2\,2}(k_1-q)+
\delta_{\vec{k}_1\,\,\vec{k}_2+\vec{Q}}\gs_{2\,1}(k_1)\gs_{2\,1}(k_1-q)
\end{pmatrix} \;.
\end{split}
\end{equation}
Here the $\gs$ are 
spin independent Green's functions (i.~e.~$\g$ (Eq.~(\ref{13})) with spin 
index $\sigma$ set equal to $+1$).
The interaction vertex in this representation is given by
\begin{equation}\label{21}
\begin{split}
\overl{\Gamma}_{a\,b}^{{\scriptscriptstyle 0}}
(\vec{k}_1,\vec{k}_2;\vec{q}+n\,\vec{Q},\vec{q}+n^\prime\vec{Q})=\frac{1}{
\beta N}\{&
\begin{pmatrix}1&0&0&0\\0&-1&0&0\\0&0&-1&0\\0&0&0&-1\end{pmatrix}\otimes
\begin{pmatrix}1&0\\0&1\end{pmatrix}U\\
&+\begin{pmatrix}2&0&0&0\\0&0&0&0\\0&0&0&0\\0&0&0&0\end{pmatrix}\otimes[
\begin{pmatrix}1&0\\0&0\end{pmatrix}V^{\scriptscriptstyle W}(\vec{k}_2-
\vec{q},\vec{k}_1,\vec{q})\\
&\qquad\qquad\qquad\qquad+\begin{pmatrix}0&0\\0&-1\end{pmatrix}
V^{\scriptscriptstyle W}(\vec{k}_2-\vec{q},\vec{k}_1,\vec{q}+\vec{Q})]\\
&-\begin{pmatrix}1&0&0&0\\0&1&0&0\\0&0&1&0\\0&0&0&1\end{pmatrix}\otimes
\begin{pmatrix}1&0\\0&-1\end{pmatrix}
V^{\scriptscriptstyle W}(\vec{k}_2-\vec{q},\vec{k}_1,\vec{k}_1-\vec{k}_2)
\,\}.
\end{split}
\end{equation}
\changed{It is written as a direct product of spin- and $(n,n^\prime)$-matrices 
with the $W$-dependent
interaction $V^{\scriptscriptstyle W}$ given by}
\begin{equation}\label{22}
\begin{split}
V^{\scriptscriptstyle W}(\vec{k},\vec{k}^\prime,\vec{q})=
-8W[&(\cos k_x^\prime\pm\cos k_y^\prime)
(\cos k_x\pm\cos k_y)\\
+&(\cos (k_x^\prime-q_x)\pm\cos (k_y^\prime-q_y))(\cos k_x\pm\cos k_y)\\
+&(\cos (k_x+q_x)\pm\cos (k_y+q_y))(\cos k_x^\prime\pm\cos k_y^\prime)\\
+&(\cos (k_x+q_x)\pm\cos (k_y+q_y))(\cos (k_x^\prime-q_x)
\pm\cos (k_y^\prime-q_y))].
\end{split}
\end{equation}

\subsection{Bethe-Salpeter Equation}

With help of the HF two-particle propagator $\overl{L}$ and the 
interaction vertex $\overl{\Gamma}$,
 we can now write down the Bethe-Salpeter equation in the form:
\begin{equation}\label{23}
\begin{split}
\overl{L}_{a\,b}(k_1,k_2;\vec{q},\vec{q}^\prime,\nu)=&
\overl{L}_{a\,b}^{{\scriptscriptstyle 0}}(k_1,k_2;\vec{q},\vec{q}^\prime,
\nu)\\
&\qquad+\sum_{\substack{k_3,k_4\\\vec{q}^{\prime\prime},\vec{q}^{\prime
\prime\prime}\\c,d}}
\overl{L}_{a\,c}^{{\scriptscriptstyle 0}}(k_1,k_3;\vec{q},\vec{q}^{\prime
\prime},\nu)\,
\overl{\Gamma}_{c\,d}^{{\scriptscriptstyle 0}}(\vec{k}_3,\vec{k}_4;
\vec{q}^{\prime\prime},
\vec{q}^{\prime\prime\prime})\,
\overl{L}_{d\,b}(k_4,k_2;\vec{q}^{\prime\prime\prime},\vec{q}^\prime,\nu).
\end{split}
\end{equation}
Unlike for the standard ($W=0$) Hubbard model,
this equation cannot be easily inverted
due to the complicated space and spin structure of the $W$ term. For
this 
reason, we apply a method due to  Hanke and Sham 
\cite{hash}, which is based on the partial transformation of the 
interaction
vertex back into
real space.
For short--range interactions, this yields finite (small-sized) matrices 
in real space.
With the Fourier-transformed correlation function and interaction vertex
\begin{align}
\widehat{L}_{a\,b}(\vec{R}_1,\vec{R}_2;\vec{q},\vec{q}^\prime,\nu)&=
-\frac{1}{\beta N}
\sum_{\substack{\vec{k}_1,\vec{k}_2\\\omega_1,\omega_2}}e^{\,i\vec{k}_1
\vec{R}_1}e^{-i\vec{k}_2\vec{R}_2}
\,\overl{L}_{a\,b}(k_1,k_2;\vec{q},\vec{q}^\prime,\nu),\\
\widehat{\Gamma}_{a\,b}^{{\scriptscriptstyle 0}}(\vec{R}_1,\vec{R}_2;
\vec{q},\vec{q}^\prime)&=
-\frac{\beta}{N}\sum_{\vec{k}_1,\vec{k}_2}e^{\,i\vec{k}_1\vec{R}_1}
e^{-i\vec{k}_2\vec{R}_2}
\,\overl{\Gamma}_{a\,b}^{{\scriptscriptstyle 0}}(\vec{k}_1,\vec{k}_2;
\vec{q},\vec{q}^\prime),
\end{align}
we obtain the Bethe-Salpeter equation in matrix form:
\begin{equation}\label{24}\begin{split}
\widehat{L}_{a\,b}(\vec{R}_1,\vec{R}_2;\vec{q},\vec{q}^\prime,\nu)=&
\widehat{L}_{a\,b}^{{\scriptscriptstyle 0}}(\vec{R}_1,\vec{R}_2;\vec{q},
\vec{q}^\prime,\nu)\\&+
\widehat{L}_{a\,c}^{{\scriptscriptstyle 0}}(\vec{R}_1,\vec{R}_3;\vec{q},
\vec{q}^{\prime\prime},\nu)
\widehat{\Gamma}_{c\,d}^{{\scriptscriptstyle 0}}(\vec{R}_3,\vec{R}_4;
\vec{q}^{\prime\prime},\vec{q}^{\prime\prime\prime})
\widehat{L}_{d\,b}(\vec{R}_4,\vec{R}_2;\vec{q}^{\prime\prime\prime},
\vec{q}^\prime,\nu).
\end{split}\end{equation}
This is now a simple matrix equation which can easily be inverted for the 
interacting two-particle propagator
$\widehat{L}$:
\begin{equation}\label{25}
\widehat{L}=\left(1-\widehat{L}^{{\scriptscriptstyle 0}}
\widehat{\Gamma}^{{\scriptscriptstyle 0}}\right)^{-1}
\widehat{L}^{{\scriptscriptstyle 0}}.
\end{equation}
In contrast to the Hubbard model we have to deal with complex 
$26\times26$ matrices for the
transverse spin-channel and with complex $52\times52$ matrices for the 
coupled charge-/ longitudinal
spin-channel.
 The RPA susceptibilities can be constructed by taking the 
$(\vec{0},\vec{0})$-matrix element in
real space, i.~e.: 
\begin{equation}\label{26}
\chi_{a\,b}(\vec{q},\vec{q}^\prime;\nu)=\widehat{L}_{a\,b}
(\vec{0},\vec{0};\vec{q},\vec{q}^\prime,\nu) 
\;.
\end{equation}
From this one obtains the retarded susceptibilities by the analytic 
continuation $i\nu\longrightarrow\omega+i\eta$.

\begin{figure}[h]
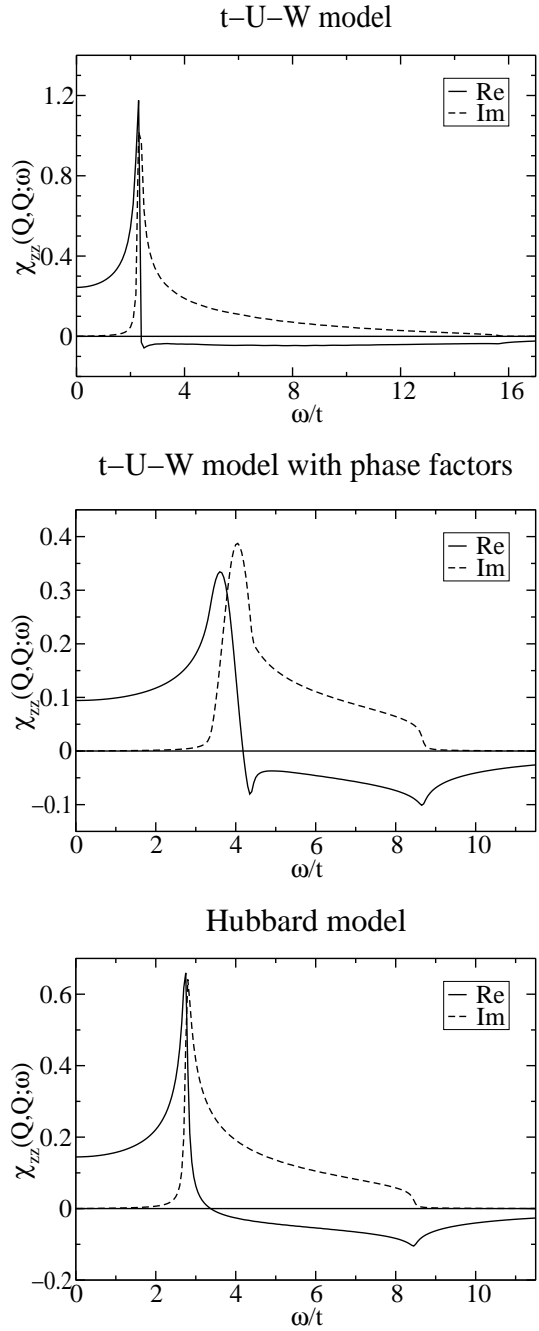

\begin{center}
\epsfig{file=fig7a.eps,width=7cm}
\end{center}
\begin{center}
\epsfig{file=fig7b.eps,width=7cm}
\end{center}
\begin{center}
\epsfig{file=fig7c.eps,width=7cm}
\end{center}
\caption[]{Longitudinal spin susceptibility 
$\chi_{\scriptscriptstyle{z\,z}}(\vec{Q},\vec{Q};\omega)$ 
of the simple $t-U-W$ model (top, $W=0.1t$),
the $t-U-W$ model with phase factors (middle, $W=0.05t$)
and the Hubbard model (bottom, $W=0t$); (as usual: $U=4t$, $T=0K$, 
$\mu=0\,t$).}
\label{fig7a}
\label{fig7b}
\label{fig7c}
\end{figure}

Starting from the idea that spin fluctuations are responsible for the 
pairing of the quasi-particles,
it is reasonable to first concentrate on the dynamic spin susceptibilities
for the antiferromagnetic nesting vector $\vec{Q}=(\pi,\pi)$ 
as a function of $\omega$, as we expect  the strongest response there.
In Fig.~\ref{fig7a} (top), $\chi_{z\,z}$ is plotted for the simple $t-U-W$
model with $W=0.1t$,
while in Fig.~\ref{fig7b} (middle) $\chi_{z\,z}$ is displayed for the  
model with phase
factors and $W=0.05t$. Both calculations can be compared with the result 
for 
the Hubbard model, reported in 
 Fig.~\ref{fig7c} (bottom).

One can clearly see that the spectral weight is mainly concentrated at 
low frequencies and 
that it is abruptly decreasing at a frequency $\omega\approx 
2\Delta_{{\scriptscriptstyle AF}}$,
which corresponds to the antiferromagnetic gap. This behavior is most 
evident
in the simple $t-U-W$ model. Moreover, one can recognize
that the overall magnitude of the longitudinal spin susceptibility is 
biggest in the simple $t-U-W$ model
and smallest in the $t-U-W$ model with phase factors.

\subsection{Effective Interaction}

\begin{figure}[t]
\begin{center}
\epsfig{file=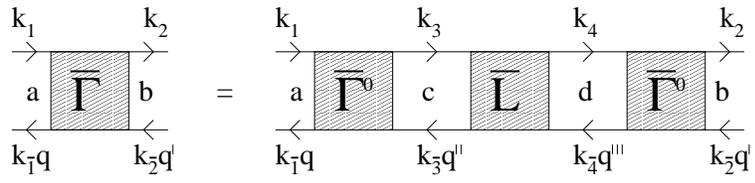,width=10cm}
\end{center}
\caption[fluctuation vertex]{Diagrammatic representation of the 
fluctuation vertex}
\label{fig8}
\end{figure}

In this section, we calculate the effective 2-particle interaction
mediated by the 
collective charge and spin
fluctuations evaluated in the preceding section.
Here, we restrict to the model without phase factor, since this is the 
only one which, according to QMC calculations, displays $d$-wave 
superconductivity. 
As a first step, we evaluate the fluctuation vertex,
from which 
we can determine
the modifications
of the bare 2-particle interaction given by the $t-U-W$ Hamiltonian.
 The calculation
of the fluctuation vertex is performed with the same techniques that were 
used to calculate the Bethe-Salpeter
equation. This gives the expression:

\begin{equation}\label{27}\begin{split}
\overl{\Gamma}_{a\,b}(\vec{k}_1,\vec{k}_2;\vec{q},\vec{q}^\prime,\nu)&=
\sum_{\substack{k_3,k_4\\\vec{q}^{\prime\prime},\vec{q}^{\prime\prime
\prime}\\c,d}}
\overl{\Gamma}_{a\,c}^{{\scriptscriptstyle 0}}(\vec{k}_1,\vec{k}_3;
\vec{q},\vec{q}^{\prime\prime})\,
\overl{L}_{c\,d}(k_3,k_4;\vec{q}^{\prime\prime},\vec{q}^{\prime\prime
\prime},\nu)\,
\overl{\Gamma}_{d\,b}^{{\scriptscriptstyle 0}}(\vec{k}_4,\vec{k}_2;
\vec{q}^{\prime\prime\prime},\vec{q}^\prime)\\
&=
\sum_{\substack{\vec{R}_1,\vec{R}_2\\\vec{R}_3,\vec{R}_4}}\,
\sum_{\substack{\vec{q}^{\prime\prime},\vec{q}^{\prime\prime\prime}\\c,d}}
\left({\textstyle\frac{-1}{\beta N}}\right)
e^{-i\vec{k}_1\vec{R}_1}
\widehat{\Gamma}_{a\,c}^{{\scriptscriptstyle 0}}(\vec{R}_1,\vec{R}_3;
\vec{q},\vec{q}^{\prime\prime})
\\&\qquad\qquad\qquad\qquad\cdot
\widehat{L}_{c\,d}(\vec{R}_3,\vec{R}_4;\vec{q}^{\prime\prime},
\vec{q}^{\prime\prime\prime},\nu)
\widehat{\Gamma}_{d\,b}^{{\scriptscriptstyle 0}}(\vec{R}_4,\vec{R}_2;
\vec{q}^{\prime\prime\prime},\vec{q}^\prime)
e^{i\vec{k}_2\vec{R}_2},
\end{split}\end{equation}
which is diagrammatically represented in Fig.~\ref{fig8}.

\begin{figure}[t]
\begin{center}
\epsfig{file=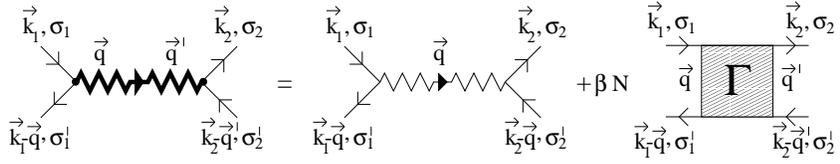,width=11cm}
\end{center}
\caption[effective interaction]{Diagrammatic representation of the 
effective interaction}
\label{fig9}
\end{figure}

Next, we have to change from the charge/spin channel representation back 
to the simple spin representation by inverting
the transformation given by Eq.~(\ref{15}):
\begin{equation}\label{28}\begin{split}
\Gamma\,^{\scriptscriptstyle\substack{\sigma_1\,\sigma_2\\
\sigma_1^\prime\,\sigma_2^\prime}}
(\vec{k}_1,\vec{k}_2;\vec{q},\vec{q}^\prime,\nu)&=
U_{\substack{\sigma_1 \\ \sigma_1^\prime}\,a}\,
\overl{\Gamma}_{a\,b}(\vec{k}_1,\vec{k}_2;\vec{q},\vec{q}^\prime,\nu)\,
(U_{\substack{\sigma_2 \\ \sigma_2^\prime}\,b})^\dagger\\&=\frac{1}{2}(
\overl{\Gamma}_{0\,0}(\vec{k}_1,\vec{k}_2;\vec{q},\vec{q}^\prime,\nu)
\sigma^0_{\sigma_1\,\sigma_1^\prime}\sigma^0_{\sigma_2^\prime\,\sigma_2}+
\overl{\Gamma}_{0\,z}(\vec{k}_1,\vec{k}_2;\vec{q},\vec{q}^\prime,\nu)
\sigma^0_{\sigma_1\,\sigma_1^\prime}\sigma^z_{\sigma_2^\prime\,\sigma_2}
\\&\quad\,\,\,+
\overl{\Gamma}_{z\,0}(\vec{k}_1,\vec{k}_2;\vec{q},\vec{q}^\prime,\nu)
\sigma^z_{\sigma_1\,\sigma_1^\prime}\sigma^0_{\sigma_2^\prime\,\sigma_2}+
\overl{\Gamma}_{z\,z}(\vec{k}_1,\vec{k}_2;\vec{q},\vec{q}^\prime,\nu)
\sigma^z_{\sigma_1\,\sigma_1^\prime}\sigma^z_{\sigma_2^\prime\,\sigma_2}
\\&\quad\,\,\,+
\overl{\Gamma}_{+\,-}(\vec{k}_1,\vec{k}_2;\vec{q},\vec{q}^\prime,\nu)
\widetilde{\sigma}^-_{\sigma_1\,\sigma_1^\prime}\widetilde{\sigma}^+_{
\sigma_2^\prime\,\sigma_2}+
\overl{\Gamma}_{-\,+}(\vec{k}_1,\vec{k}_2;\vec{q},\vec{q}^\prime,\nu)
\widetilde{\sigma}^+_{\sigma_1\,\sigma_1^\prime}\widetilde{\sigma}^-_{
\sigma_2^\prime\,\sigma_2}).
\end{split}\end{equation}
Since we want to use the effective interaction in order to write 
an effective Hamiltonian, only the static limit
of the fluctuation vertex has to be considered.
Thus, simple diagrammatic rules \cite{fewa}  yield for the correction to 
the bare
interaction:
\begin{equation}\label{29}
\widetilde{V}\,^{\scriptscriptstyle\substack{\sigma_1\,\sigma_2\\
\sigma_1^\prime\,\sigma_2^\prime}}
(\vec{k}_1,\vec{k}_2;\vec{q},\vec{q}^\prime)=
\Gamma\,^{\scriptscriptstyle\substack{\sigma_1\,\sigma_2\\
\sigma_1^\prime\,\sigma_2^\prime}}
(\vec{k}_1,\vec{k}_2;\vec{q},\vec{q}^\prime,0)\cdot(-\beta N)\cdot(-1).
\end{equation}
The effective interaction can then be written as (see Fig.~\ref{fig9} 
for diagrammatic representation)
\begin{equation}\label{30}
V_{eff}\,^{\scriptscriptstyle\substack{\sigma_1\,\sigma_2\\
\sigma_1^\prime\,\sigma_2^\prime}}
(\vec{k}_1,\vec{k}_2;\vec{q},\vec{q}^\prime)=
V\,^{\scriptscriptstyle\substack{\sigma_1\,\sigma_2\\\sigma_1^\prime\,
\sigma_2^\prime}}
(\vec{k}_1,\vec{k}_2;\vec{q})+
\widetilde{V}\,^{\scriptscriptstyle\substack{\sigma_1\,\sigma_2\\
\sigma_1^\prime\,\sigma_2^\prime}}
(\vec{k}_1,\vec{k}_2;\vec{q},\vec{q}^\prime)
\end{equation}
with the bare 2-particle interaction 
\begin{equation}\label{31}
V\,^{\scriptscriptstyle\substack{\sigma_1\,\sigma_2\\\sigma_1^\prime\,
\sigma_2^\prime}}
(\vec{k}_1,\vec{k}_2;\vec{q})=U\delta_{\sigma_1\,\sigma_1^\prime}
\delta_{\sigma_2\,\sigma_2^\prime}
\delta_{\sigma_1\,\bar{\sigma}_2}+
V^{\scriptscriptstyle W}(\vec{k}_2-\vec{q},\vec{k}_1;\vec{q})
\delta_{\sigma_1\,\sigma_1^\prime}\delta_{\sigma_2\,\sigma_2^\prime}
\end{equation}
and $V^{\scriptscriptstyle W}$ given by Eq.~(\ref{22}). 
Since one has to consider the pairing of the Hartree-Fock
quasiparticles, 
the effective interaction has to be transformed into 
 the $\gamma$-base,
which produces additional coherence 
factors \cite{swz}.

For physical reasons, only the pairing of particles with opposite spin 
(singlet-pairing) was considered.
In order to take into account  the effect of the dynamics 
on top of our static approximation,
we follow Ref.~\onlinecite{swz}, and introduce a cutoff frequency
$\omega_{\scriptstyle c}$,
 analogous to the Debye frequency $\omega_{\scriptscriptstyle D}$
in the standard BCS theory. This ensures
that only particles within an interval of width $\hbar
\omega_{\scriptstyle c}$ above and
below the Fermi energy $E_{\scriptscriptstyle F}$ are paired.

The motivation for this cutoff frequency $\omega_{\scriptstyle c}$ 
becomes clear if one looks at the 
spin susceptibilities of the simple $t-U-W$ model in Fig.~\ref{fig7a}. We 
have already shown in the
preceding section that the spectral weight is concentrated at low 
frequencies.
Under the condition that the spin fluctuations are responsible for the 
pairing of the quasi-particles,
the longitudinal spin susceptibility gives quite naturally a cutoff 
frequency of the size of the
antiferromagnetic gap ($\omega_c\approx 
2\Delta_{{\scriptscriptstyle AF}}$).

This implies that for hole dopings away from half-filling, only the 
intra-valence band matrix elements have to
be considered. The pairing part of the effective Hamiltonian can thus 
formally be written as
\begin{equation}\label{32}
{\mathcal{H}}^{\scriptscriptstyle pair}=
\frac{1}{2N}\sideset{}{'}\sum_{\substack{\vec{k},\vec{k}^\prime\\
\sigma,\sigma^\prime}}
V^{\scriptscriptstyle pair}_{\sigma\,\sigma^\prime}(\vec{k},\vec{k}^\prime)
\Theta(\omega_{\scriptstyle c}-|E^v(\vec{k})-E_{\scriptscriptstyle F}|)
\Theta(\omega_{\scriptstyle c}-|E^v(\vec{k}^\prime)
-E_{\scriptscriptstyle F}|)
\gvk{\vec{k}^\prime,\sigma^\prime}\gvk{-\vec{k}^\prime,-\sigma^\prime}
\gv{-\vec{k},-\sigma}\gv{\vec{k},\sigma},
\end{equation}
where $E^v(\vec{k})=-E(\vec{k})$ is the valance band energy.

\begin{figure}[h]
\begin{center}
\epsfig{file=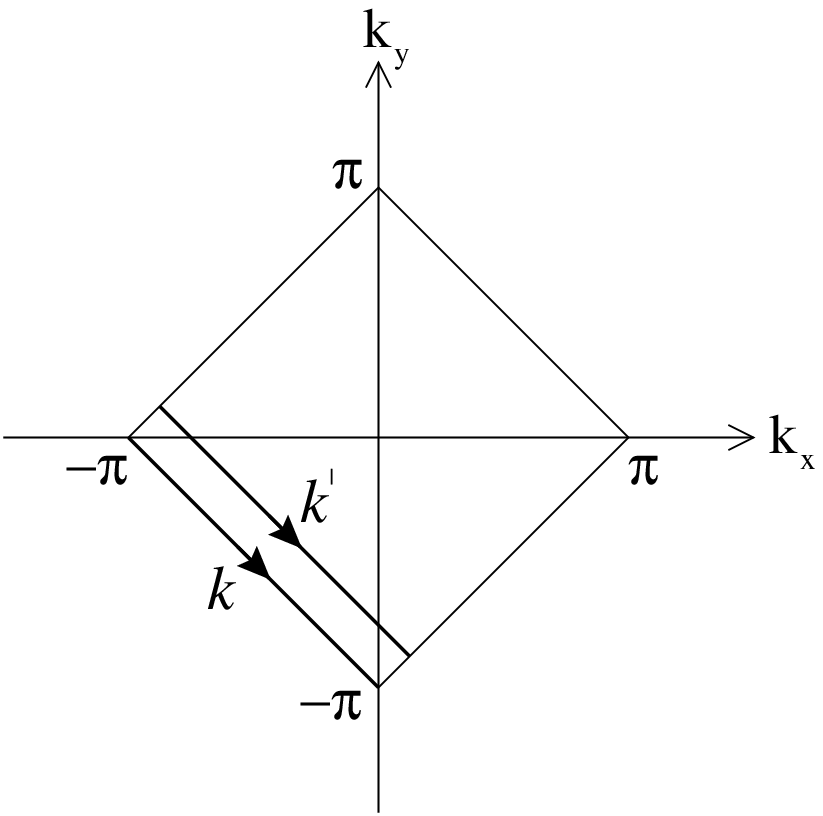,width=3.5cm}
\end{center}
\begin{center}
\epsfig{file=fig10b.eps,width=8cm}
\end{center}
\begin{center}
\epsfig{file=fig10c.eps,width=8cm}
\end{center}
\caption[]{Direct part $V^{\scriptscriptstyle pair}_{\sigma\,\sigma}
(\vec{k},\vec{k}^\prime)$ and
exchange part $-V^{\scriptscriptstyle pair}_{\sigma\,-\sigma}
(\vec{k},-\vec{k}^\prime)$ of
the pairing-potential for the simple $t-U-W$ model (top, $W=0.1t$)
and for the Hubbard model (bottom, $W=0t$).
The paths shown above were followed by $\vec{k}$ and $\vec{k}^\prime$ 
through the MBZ.
The distance between $\vec{k}$ and $\vec{k}^\prime$ amounts 
$\delta\!\vec{k}\approx0.14\pi$
($U=4t$, $T=0K$, $\mu=0\,t$, lattice size: $100\times100$).}
\label{fig10b}
\label{fig10c}
\end{figure}

\begin{figure}[h]
\begin{center}
\epsfig{file=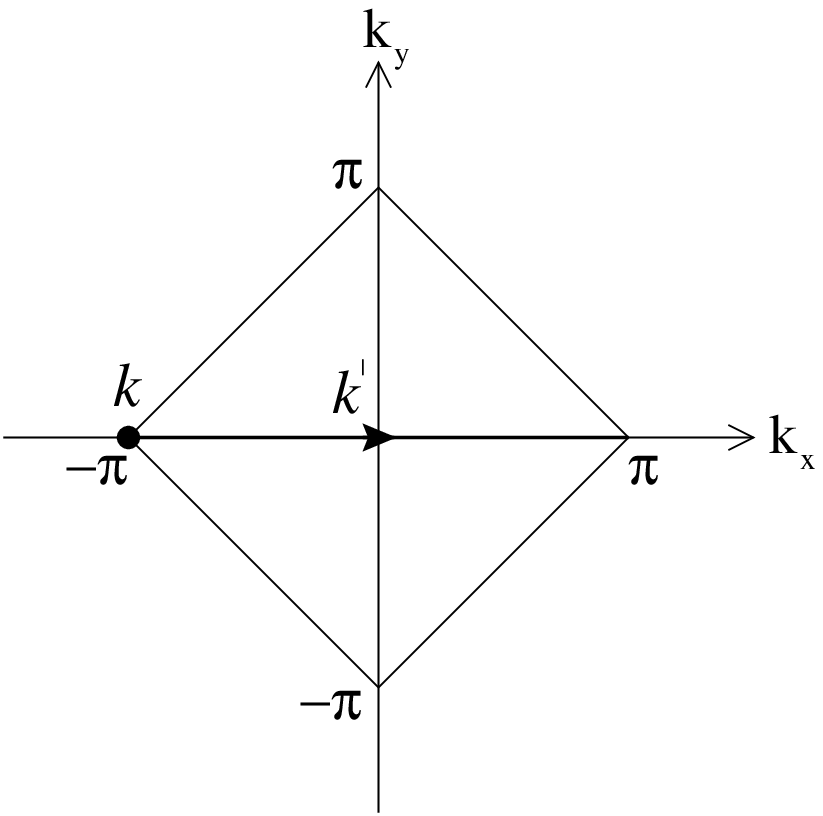,width=3.5cm}
\end{center}
\begin{center}
\epsfig{file=fig11b.eps,width=8cm}
\end{center}
\begin{center}
\epsfig{file=fig11c.eps,width=8cm}
\end{center}
\caption[]{Direct part $V^{\scriptscriptstyle pair}_{\sigma\,\sigma}
(\vec{k},\vec{k}^\prime)$ and
exchange part $-V^{\scriptscriptstyle pair}_{\sigma\,-\sigma}
(\vec{k},-\vec{k}^\prime)$ of
the pairing-potential for the simple $t-U-W$ model (top, $W=0.1t$)
and for the Hubbard model (bottom, $W=0t$).
The path shown above was followed by $\vec{k}^\prime$ through the MBZ with
$\vec{k}$ held constant ($U=4t$, $T=0K$, $\mu=0\,t$, lattice size: 
$100\times100$).}
\label{fig11b}
\label{fig11c}
\end{figure}

\begin{figure}[h]
\begin{center}
\epsfig{file=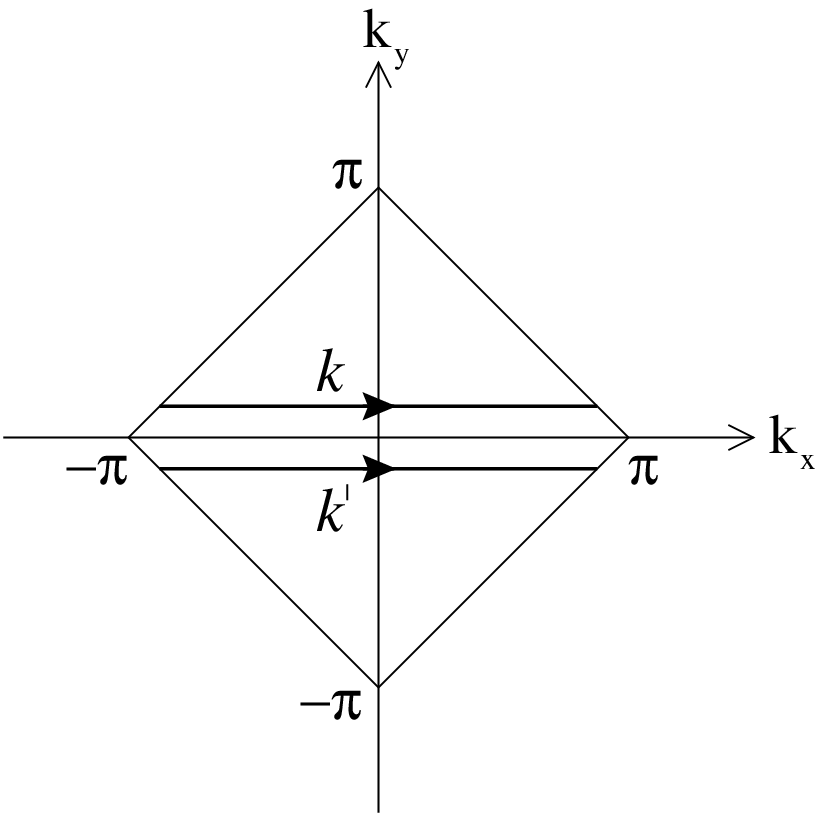,width=3.5cm}
\end{center}
\begin{center}
\epsfig{file=fig12b.eps,width=8cm}
\end{center}
\begin{center}
\epsfig{file=fig12c.eps,width=8cm}
\end{center}
\caption[]{Direct part $V^{\scriptscriptstyle pair}_{\sigma\,\sigma}
(\vec{k},\vec{k}^\prime)$ and
exchange part $-V^{\scriptscriptstyle pair}_{\sigma\,-\sigma}
(\vec{k},-\vec{k}^\prime)$ of
the pairing-potential for the simple $t-U-W$ model (top, $W=0.1t$)
and for the Hubbard model (bottom, $W=0t$)
The paths shown above were followed by $\vec{k}$ and $\vec{k}^\prime$ 
through the MBZ.
The distance between $\vec{k}$ and $\vec{k}^\prime$ amounts 
$\delta\!\vec{k}\approx0.2\pi$
($U=4t$, $T=0K$, $\mu=0\,t$, lattice size: $100\times100$).}
\label{fig12b}
\label{fig12c}
\end{figure}

The direct interaction $V^{\scriptscriptstyle pair}_{\sigma\,\sigma}
(\vec{k},\vec{k}^\prime)$, which is given by 
$\sigma=\sigma^\prime$ spin indices contains, besides the longitudinal 
spin fluctuations, also the bare interactions
and the charge fluctuations. The exchange interaction with 
$\sigma=-\sigma^\prime$ consists of
the transverse spin fluctuations only. In Figs.~\ref{fig10b} to \ref{fig12c},
the direct interaction and the exchange interaction were plotted 
for different paths of $\vec{k}$ and $\vec{k}^\prime$ through the 
magnetic Brillouin zone (MBZ)
for the simple $t-U-W$ model and the Hubbard model, respectively.
The exchange interaction was plotted there as 
$-V^{\scriptscriptstyle pair}_{\sigma\,-\sigma}(\vec{k},-\vec{k}^\prime)$,
since it has exactly this form, with negative sign, in the BCS gap 
equation (see Eq.~(\ref{34})).

Comparing the graphs for the simple $t-U-W$ model ($W=0.1t$) with the simple 
Hubbard model, one can easily see
that the $W$ term amplifies the attractive parts of the direct 
interaction, whereas the attractive parts of the 
exchange interaction remain constant (see e.~g.~Fig.~\ref{fig11b}). 
However, the repulsive parts of the direct interaction and
the exchange interaction were both attenuated considerably by increasing 
$W$ (see e.~g.~Fig.~\ref{fig12b}).
Therefore,  the pairing of the quasi-particles is favored in the simple
$t-U-W$ model
altogether.

\section{BCS Gap Equation}
\label{bcs}

Finally, we want to solve the BCS gap equation for the effective pairing 
interaction.
Starting point is the effective Hamiltonian, as obtained in the
previous Section 
\begin{equation}\label{33}\begin{split}
{\mathcal{H}}_{eff}=&
\sideset{}{'}\sum_{\vec{k},\sigma}
(E^v(\vec{k})-\mu)\gvk{\vec{k},\sigma}\gv{\vec{k},\sigma}\\
&+\frac{1}{2N}\sideset{}{'}\sum_{\substack{\vec{k},\vec{k}^\prime\\
\sigma,\sigma^\prime}}
V^{\scriptscriptstyle pair}_{\sigma\,\sigma^\prime}(\vec{k},\vec{k}^\prime)
\Theta(\omega_{\scriptstyle c}-|E^v(\vec{k})-E_{\scriptscriptstyle F}|)
\Theta(\omega_{\scriptstyle c}-|E^v(\vec{k}^\prime)-E_{\scriptscriptstyle 
F}|)
\gvk{\vec{k}^\prime,\sigma^\prime}\gvk{-\vec{k}^\prime,-\sigma^\prime}
\gv{-\vec{k},-\sigma}\gv{\vec{k},\sigma}.
\end{split}\end{equation}
With this Hamiltonian we want to study the superconducting properties of 
the simple
$t-U-W$ model for different hole doping and different values of the model 
parameter $W$.
The BCS gap equation becomes
\begin{equation}\label{34}
\Delta(\vec{k})=-\frac{1}{N}\sideset{}{'}\sum_{\vec{k}^\prime}
(V_{\up\,\up}(\vec{k},\vec{k}^\prime)
-V_{\up\,\dn}(\vec{k},-\vec{k}^\prime))
\frac{\Delta(\vec{k}^\prime)}{2E(\vec{k}^\prime)}.
\end{equation}
Here we used the abbreviations
\begin{align}
E(\vec{k})&=\sqrt{\xi^2(\vec{k})+\Delta^2(\vec{k})},\\
\xi(\vec{k})&=E^v(\vec{k})-\mu,\\
V_{\sigma\,\sigma^\prime}(\vec{k},\vec{k}^\prime)&=
V^{\scriptscriptstyle pair}_{\sigma\,\sigma^\prime}(\vec{k},\vec{k}^\prime)
\Theta(\omega_{\scriptscriptstyle A}-|E^v(\vec{k})
-E_{\scriptscriptstyle F}|)
\Theta(\omega_{\scriptscriptstyle A}-|E^v(\vec{k}^\prime)
-E_{\scriptscriptstyle F}|).
\end{align}
The gap equation (\ref{34}) was iterated by assuming  different 
symmetries of the superconducting 
order parameter, however,  only  $d$-wave solutions turn out to converge.

\begin{figure}[h]
\begin{center}
\epsfig{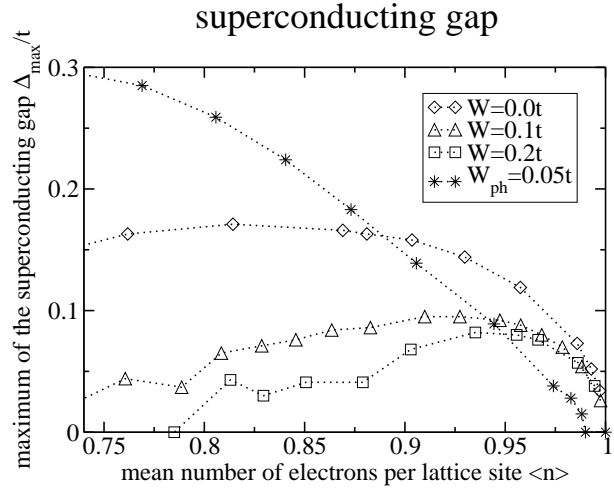}
\end{center}
\caption[]{Maximum value of the superconducting gap function 
$\Delta_{\scriptscriptstyle max}=\max \Delta(\vec{k})$
as a function of the mean electron number per lattice site 
$\langle n\rangle$ for different values
of the model parameter $W$ ($U=4t$, $T=0K$, lattice size: 
$20\times20$, $d$-wave symmetry). \changed{The curves for the simple $t-U-W$ model 
are depicted as diamonds, triangles and squares and the curve for the 
$t-U-W$ model with phase factors is depicted as stars.}}
\label{fig13}
\end{figure}

\changed{
The results for the simple $t-U-W$ model are shown in Fig.~\ref{fig13}
as diamonds, triangles and squares. 
For all values of $W$ considered,
 the superconducting gap becomes zero at half filling, which is consistent
with  QMC results. 
 It can be shown
easily with the aid of equation (\ref{34}), that even including 
interband
matrix elements doesn't change this result.
 This is due to the fact that the
band
gap is still quite large for these values of $W$ (cf. Fig.~\ref{fig3a})
for interband matrix elements to contribute substantially to the gap equation.
On the other hand, all curves seem to indicate that the
superconducting phase starts at very small doping, which  is in contrast
with experiments.
While there are no conclusive QMC results about the $t-U-W$ model in
this region, due to the minus sign problem,
 the limitations of our perturbative procedure applied for moderate
values of $U$ suggest to consider this result with due care.
Indeed, strong {\it phase} fluctuations, not included in a
 BCS-type of calculation like Eq.(\ref{34})
are known to be  important and to 
suppress superconductivity at small doping.
}

\changed{
Figure \ref{fig13} also shows, that the optimal hole doping for the 
simple $t-U-W$ model is moving closer to half-filling with increasing $W$.
For comparison, we show in Fig.~\ref{fig13} (stars)
a result for the $t-U-W$ model
with phase factor.
 One can see, that the superconducting gap is
strongly suppressed for hole doping near half-filling, compared to
the simple $t-U-W$ (and Hubbard)  model.
The effect is here even stronger than for the simple $t-U-W$ model,
since in
 the model with phase factors the gap {\it increases} with $W$.
}

\changed{
It remains the question as to why the superconducting gap is getting smaller 
with increasing $W$ in the simple $t-U-W$ model, in spite 
of the fact that the bare attraction between the quasi-particles is 
enhanced by the $W$ term.}
To understand this, one has to keep in mind that there are other 
important quantities,
like the cutoff frequency or the density of states at the Fermi level, 
which have an important effect on
the magnitude of the superconducting gap. How the superconducting gap
depends on these quantities  is qualitatively seen already
in the weak-coupling solution of the original BCS equation for an 
attractive $\delta$-potential \cite{fewa}:
\begin{equation}\label{35}
\Delta=2\hbar\omega_{\scriptscriptstyle D}
\exp\left(-\frac{1}{N(0)g}\right).
\end{equation}
A stronger coupling $g$ is increasing the superconducting gap, while a 
smaller cutoff frequency 
$\omega_{\scriptscriptstyle D}$ is decreasing it.
Moreover, if the density of states at the Fermi level $N(0)$ is
reduced, the superconducting gap is  getting smaller, as well.
This final point is decisive, since the $W$ term is broadening the energy 
bands considerably, thus reducing
the density of states dramatically.
Therefore,  the fact that the superconducting gap is getting smaller with 
increasing $W$
can be understood within our approximation.

\section{Comparison with QMC Results and Conclusions}
\label{summary}

In this paper, we have first shown that the standard Hartree-Fock 
approximation can describe the 
antiferromagnetic properties of the $t-U-W$ model, especially the 
single-particle energy bands, in
surprisingly good agreement with the QMC simulations. On the other
hand, like for the Hubbard model~\cite{swz}, it is unable 
to reproduce the transition to a $d$-wave superconductor observed in the 
simple $t-U-W$ model
in numerical simulations (QMC). 
\changede{
In order to partly overcome 
this short-coming 
we adopt a
time-dependent HF
or generalized RPA calculations, which we
present in the second part of our paper.
}

The standard Hartree-Fock approximation captures only 
the ``high-energy'' physics, and is thus capable of reproducing band-widths
and the overall features of the single-particle spectral-function. 
Equivalently, short-range pairing-correlation functions should 
be well reproduced within this approximation. 
This is indeed the case. At short length 
scales (as shown in table 1) the extended
s-wave vertex contribution to the pairing correlation functions is dominant
in QMC simulations of the simple t-U-W model. It is only at {\it larger} 
distances that the d-wave 
pairing correlations become dominant. This crossover 
from {\it short range} to
{\it long range} properties is not reproduced within the standard 
mean-field 
approximation. 
Alternatively, for the t-U-W model with
phase factors  d-wave pairing dominates in QMC simulations at small 
length scales (see table 2).

On the other hand, by including charge and spin fluctuations within a 
time-dependent HF, or generalized RPA summation of ladder
and bubble diagrams, we were able to obtain an effective
attraction between the quasi-particles, which is
  enhanced by $W$. Moreover we obtain a
corresponding superconducting order parameter with the correct d-wave 
symmetry.

\begin{figure}[h]
\begin{center}
\epsfig{file=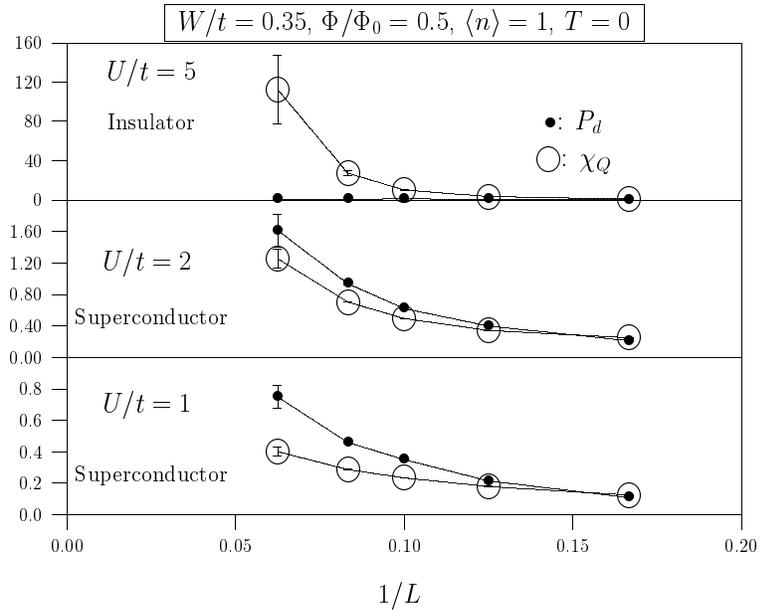,width=10cm}
\end{center}
\caption[]{Vertex contribution to the $d$-wave pairing correlations  
($\bullet$)
and staggered spin 
susceptibility  ($\bigcirc$)
for the simple $t-U-W$ model, obtained from QMC simulations.}
\label{fig14}
\end{figure}

The QMC results indeed show that  as $W$ increases at fixed $U$ or  
$U$ decreases at fixed $W$, an instability towards $d$-wave 
superconductivity 
occurs.  To illustrate this, Fig.~\ref{fig14} plots the vertex contribution 
to the $d$-wave pairing correlations as well as the  staggered spin 
susceptibility. As apparent at low $U$ for fixed $W$,  the 
superconducting $d$-wave becomes the
 leading 
instability.
\changede{
Finally, it should be pointed out that our
Hartree-Fock-Bethe-Salpeter procedure
is of {\it perturbative} nature
and, therefore,
our results should be considered with due care, due to the fact that
the interaction is not small.
}

\section*{Acknowledgments}

The authors express their deep gratitude to Prof.~M.~Imada for many 
insightful 
discussions and clarifications of physical points. In this connection 
one of us
(W.~H.) is grateful to Prof.~Imada for the warm hospitality experienced 
during visits
at the ISSP in Tokyo. He acknowledges the support granted by the joint 
German - Japanese cooperation project 
(DFG -- JSPS: 446 JAP -- 113/114/0), which was 
central for the success of this work.
\changed{Partial support by the DFG-Project Ha 1537/14-1 is also acknowledged.}

\begin{appendix}

\section{Details of the HF calculations}
\label{a1}
\subsection{Antiferromagnetic Mean Field}
We start with the following Ansatz for the mean field parameters:
\begin{align}
\lk\cks{i}\cs{i}\rk&=n_1+\sigma e^{i\vec{Q}\cdot\vec{i}}n_2,\\
\lk\cks{i} f(\vec{\delta})\c{\vec{i}+\vec{\delta},\sigma}\rk&=n_3,\\
\lk\ck{\vec{i}+\vec{\delta},\sigma} f(\vec{\delta})f(\vec{\delta}^\prime)
\c{\vec{i}+\vec{\delta}^\prime,\sigma}\rk&=n_4+\sigma 
e^{i\vec{Q}\cdot\vec{i}}n_5,
\end{align}
where $\vec{Q}$ denotes the antiferromagnetic nesting-vector 
$\vec{Q}=(\pi,\pi)$. 
Here the expectation values $\lk\cdots\rk$ contain also an average over 
all $\vec{\delta}$ and
$\vec{\delta}^\prime$, whenever explicitly present.
At  half filling, it is easy to show that $n_1=\frac{1}{2}$, and 
$n_4=\frac{1}{8}$. 
The mean--field Hamiltonian can then be written as
\begin{equation}\label{5}
{\mathcal{H}}^{\scriptscriptstyle MF}_{\scriptscriptstyle tUW}=
\sideset{}{'}\sum_{\vec{k},\sigma}
(\cck{\vec{k},\sigma},\cck{\vec{k}+\vec{Q},\sigma})
\begin{pmatrix} \varepsilon(\vec{k})& \sigma \Delta(\vec{k})\\
\sigma \Delta(\vec{k})& -\varepsilon(\vec{k})\end{pmatrix}
\begin{pmatrix}\cc{\vec{k},\sigma}\\\cc{\vec{k}+\vec{Q},\sigma}
\end{pmatrix}
+\widetilde{E}^{\scriptscriptstyle MF}_{\scriptscriptstyle tUW}
\end{equation}
with
\begin{align}
\varepsilon(\vec{k})&=-2t(\cos k_x+\cos k_y)-96Wn_3(\cos k_x\pm\cos k_y),
\label{5a}\\
\Delta(\vec{k})&=-Un_2+32Wn_5-8Wn_2(\cos k_x\pm\cos k_y)^2\label{5b}
\end{align}
and
\begin{equation}\label{6}
\widetilde{E}^{\scriptscriptstyle MF}_{\scriptscriptstyle tUW}=
+UNn_2^2+192WNn_3^2
-64WN(\textstyle\frac{1}{16}+n_2n_5).
\end{equation}
Here, and in the following equations, the upper sign stands for the simple
$t-U-W$ model, while the lower sign is used for the 
$t-U-W$ model with phase factors. The Hamiltonian can be diagonalized by 
the usual transformation
\begin{equation}\label{7}
\begin{pmatrix}
\gc{\vec{k},\sigma}\\\gv{\vec{k},\sigma}
\end{pmatrix}=
\begin{pmatrix}
u(\vec{k})&\sigma v(\vec{k})\\
v(\vec{k})&-\sigma u(\vec{k})
\end{pmatrix}
\begin{pmatrix}
\cc{\vec{k},\sigma}\\\cc{\vec{k}+\vec{Q},\sigma}
\end{pmatrix}
\end{equation}
with
\begin{align}
u(\vec{k})&=\left[\frac{1}{2}\left(1+\frac{\varepsilon(\vec{k})}
{E(\vec{k})}\right)\right]^\frac{1}{2},\\
v(\vec{k})&=\left[\frac{1}{2}\left(1-\frac{\varepsilon(\vec{k})}
{E(\vec{k})}\right)\right]^\frac{1}{2}
\end{align}
and
\begin{equation}\label{8}
E(\vec{k})=\sqrt{\varepsilon^2(\vec{k})+\Delta^2(\vec{k})}.
\end{equation}
The resulting Hamiltonian is given by
\begin{equation}\label{9}
{\mathcal{H}}^{\scriptscriptstyle MF}_{\scriptscriptstyle tUW}=
\sideset{}{'}\sum_{\vec{k},\sigma}
E(\vec{k})(\gck{\vec{k},\sigma}\gc{\vec{k},\sigma}-\gvk{\vec{k},\sigma}
\gv{\vec{k},\sigma})
+\widetilde{E}^{\scriptscriptstyle MF}_{\scriptscriptstyle tUW}.
\end{equation}
Already at this point, one can see from equations (\ref{5a}) and 
(\ref{8}) that the parameter $n_3$ produces extremely
wide bands in the simple $t-U-W$ model.
On the other hand, it is straightforward to see
that for the alternative $t-U-W$ model with phase factors one must have
$n_3\equiv0$, in order to preserve the symmetry of the energy bands under 
interchange of x- and y-directions. 
This explains why the bandwidth of the $t-U-W$ model with phase
factors is drastically smaller, and essentially the
same as the one of  the simple Hubbard model.

\subsection{Superconducting Mean Field}
Here, the mean field parameters are chosen as:
\begin{align}
\lk\cks{i}\cs{i}\rk&=n_1,\\
\lk\cks{i}f(\vec{\delta})\c{\vec{i}+\vec{\delta},\sigma}\rk&=n_3,\\
\lk\ck{\vec{i}+\vec{\delta},\sigma}f(\vec{\delta})f(\vec{\delta}^\prime)
\c{\vec{i}+\vec{\delta}^\prime,\sigma}\rk&=n_4,\\
\lk\ck{\vec{i},\sigma}\ck{\vec{i},-\sigma}\rk&=\sigma s_1,\\
\lk\ck{\vec{i},\sigma}f(\vec{\delta})
\ck{\vec{i}+\vec{\delta},-\sigma}\rk&=\sigma s_2,\\
\lk\ck{\vec{i}+\vec{\delta},\sigma}f(\vec{\delta})f(\vec{\delta}^\prime)
\ck{\vec{i}+\vec{\delta}^\prime,-\sigma}\rk&=\sigma s_3.
\end{align}
The superconducting parameters can be divided into two groups: $s_2$ 
stands for the nearest--neighbor singlet
pairing, which is favored by the $W$ term (Eq.~(\ref{4a})). On the other 
hand, $s_1$ and $s_3$ for $\delta=\delta^\prime$
represent the on-site singlet pairing which is suppressed by the Hubbard 
$U$.
The mean--field Hamiltonian thus is
\begin{equation}\label{10}
{\mathcal{H}}^{\scriptscriptstyle MF}_{\scriptscriptstyle tUW}=
\sum_{\vec{k}}
(\cck{\vec{k},\up},\cc{-\vec{k},\dn})
\begin{pmatrix} \varepsilon(\vec{k})& \Delta(\vec{k})\\
\Delta(\vec{k})& -\varepsilon(\vec{k})\end{pmatrix}
\begin{pmatrix}\cc{\vec{k},\up}\\\cck{-\vec{k},\dn}\end{pmatrix}
+\widetilde{E}^{\scriptscriptstyle MF}_{\scriptscriptstyle tUW}
\end{equation}
with the single--particle energy $\varepsilon(\vec{k})$ and the gap 
parameter $\Delta(\vec{k})$ given by:
\begin{align}
\varepsilon(\vec{k})&=-2t(\cos k_x+\cos k_y)\nonumber\\
&\quad -96Wn_3(\cos k_x\pm\cos k_y),\label{10a}\\
\Delta(\vec{k})&=+Us_1\nonumber\\
&\quad -8Ws_1(\cos k_x\pm\cos k_y)^2\nonumber\\
&\quad -32Ws_2(\cos k_x\pm\cos k_y)\nonumber\\
&\quad -32Ws_3.\label{10b}
\end{align}
The energy constant 
$\widetilde{E}^{\scriptscriptstyle MF}_{\scriptscriptstyle tUW}$ 
stands for:
\begin{equation}\label{11}
\widetilde{E}^{\scriptscriptstyle MF}_{\scriptscriptstyle tUW}=
-UNs_1^2+64WN(s_1s_3+s_2^2-{\textstyle\frac{1}{16}})+192WN n_3^2.
\end{equation}
Eq.~(\ref{10}) can be diagonalized with a Bogoliubov -- de Gennes 
transformation, similar to Eq. (\ref{7}), i.~e.~
\begin{align}
\begin{pmatrix}
\gc{\vec{k}}\\\gv{\vec{k}}
\end{pmatrix}&=
\begin{pmatrix}
u(\vec{k})&v(\vec{k})\\
v(\vec{k})&-u(\vec{k})
\end{pmatrix}
\begin{pmatrix}
\cc{\vec{k},\up}\\\cck{-\vec{k},\dn}
\end{pmatrix},\\
u(\vec{k})&=\left[\frac{1}{2}\left(1+\frac{\varepsilon(\vec{k})}
{E(\vec{k})}\right)\right]^\frac{1}{2},\\
v(\vec{k})&=\left[\frac{1}{2}\left(1-\frac{\varepsilon(\vec{k})}
{E(\vec{k})}\right)\right]^\frac{1}{2}
\sign(\Delta(\vec{k})).
\end{align}
The resulting Hamiltonian now becomes
\begin{equation}\label{11a}
{\mathcal{H}}^{\scriptscriptstyle MF}_{\scriptscriptstyle tUW}=
\sum_{\vec{k}}
E(\vec{k})(\gck{\vec{k}}\gc{\vec{k}}-\gvk{\vec{k}}\gv{\vec{k}})
+\widetilde{E}^{\scriptscriptstyle MF}_{\scriptscriptstyle tUW},
\end{equation}
with the usual relation (Eq.~(\ref{8})) for $E(\vec{k})$.

The paramagnetic solutions can be easily obtained by setting the 
antiferromagnetic parameters in
Eq. (\ref{5}), or the superconducting parameters in Eq. (\ref{10}), 
equal to zero.

\end{appendix}

\ifx\undefined\andword \def\andword{and} \fi 
\ifx\undefined\submittedto \def\submittedto{submitted to } \fi 
\ifx\undefined\toapperarin \def\toappearin{to appear in } \fi 


\clearpage

\section*{Figures and Tables}

\clearpage

\clearpage

\clearpage

\clearpage

\clearpage

\clearpage

\clearpage

\begin{table}[h]
\begin{center}
\begin{tabular}{|c|c|c|} \hline
$ \vec{r} $ &  $P_{s}^{v} (\vec{r}) $ & $ P_{d}^{v} (\vec{r}) $ \\ \hline
$(0,0)$  &  $ 0.2950  \pm 0.0018 $  & $ 0.1304  \pm  0.0011 $  \\
$(0,1)$  &  $ 0.0932  \pm 0.0009 $  & $ 0.0238  \pm  0.0006 $  \\
$(0,2)$  &  $ 0.0076  \pm 0.0002 $  & $ 0.0252  \pm  0.0003  $  \\
\hline
\end{tabular}
\end{center}
\caption{ Short range  vertex contribution of pair-field 
correlations  in the
extended $s$- and $d$-wave channels for the simple $t-U-W$ model, 
obtained from QMC simulations. 
Here we consider an $L = 24 $ lattice
at $W/t = 0.35 $, $ U/t =2 $ and $\langle n \rangle = 1 $.   We choose
antiperiodic (periodic) boundary conditions in the x (y) direction. 
The distance $ \vec{r} $ is in units of the lattice constant. }
\end{table}

\begin{table}[h]
\begin{center}
\begin{tabular}{|c|c|c|} \hline
$ \vec{r} $ &  $P_{s}^{v} (\vec{r}) $ & $ P_{d}^{v} (\vec{r}) $ \\ \hline
$(0,0)$  &  $ 1.661 \pm 0.079 $  & $ 3.082 \pm  0.066 $  \\
$(0,1)$  &  $ 0.373 \pm 0.020 $  & $ 1.107 \pm  0.015 $  \\
$(0,2)$  &  $ 0.091 \pm 0.008 $  & $ 0.119 \pm  0.004 $  \\
\hline
\end{tabular}
\end{center}
\caption{ Short range  vertex contribution of pair-field correlations  
in the
extended $s$- and $d$-wave channels for the $t-U-W$ model with phase 
factors, obtained from QMC simulations. Here we consider an $L = 10 $ 
lattice
at $W/t = 0.35 $, $ U/t =4 $ and $\langle n \rangle = 1 $.   We choose
periodic  boundary conditions.  The distance
$ \vec{r} $ is in units of the lattice constant. }
\end{table}

\end{document}